\documentclass[aps, pra, a4paper, amsfonts, amssymb, amsmath, reprint, showkeys, nofootinbib, twoside, superscriptaddress, longbibliography]{revtex4-1}
\usepackage[english]{babel}
\usepackage[utf8]{inputenc}
\usepackage[colorinlistoftodos, color=green!40, prependcaption]{todonotes}
\usepackage{amsthm}
\usepackage{mathtools}
\usepackage{physics}
\usepackage{xcolor}
\usepackage{graphicx}
\usepackage{siunitx}
\usepackage[T1]{fontenc}

\makeatletter
\newcommand{\pushright}[1]{\ifmeasuring@#1\else\omit\hfill$\displaystyle#1$\fi\ignorespaces}
\newcommand{\pushleft}[1]{\ifmeasuring@#1\else\omit$\displaystyle#1$\hfill\fi\ignorespaces}
\makeatother

\newcommand{\fig}[1]{Fig.~\ref{#1}}
\newcommand{\figs}[1]{Figs.~\ref{#1}}
\newcommand{\eq}[1]{Eq.~(\ref{#1})}

\usepackage[acronym,nonumberlist]{glossaries}
\newacronym{NP}{NP}{nanoparticle}
\newacronym{GAP}{GAP}{Gaussian approximation potential~\cite{bartok_2010}}
\newacronym{SOAP}{SOAP}{smooth overlap of atomic positions~\cite{bartok_2013}}
\newacronym{fcc}{fcc}{face-centered cubic}
\newacronym{NS}{NS}{nested sampling}
\newacronym{EAM}{EAM}{embedded-atom method}
\newacronym{DFT}{DFT}{density-functional theory}
\newacronym{ML}{ML}{machine learning}
\newacronym{MLP}{MLP}{{\gls{ML}} potential}
\newacronym{PES}{PES}{potential energy surface}
\newacronym{HER}{HER}{the hydrogen evolution reaction}
\newacronym{PBE}{PBE}{Perdew-Burke-Ernzerhof~\cite{perdew_1996}}
\newacronym{PBE-DFT}{PBE-DFT}{{\gls{DFT}} with the {\gls{PBE}} exchange-correlation functional}
\newacronym{MD}{MD}{molecular dynamics}
\newacronym{AIMD}{AIMD}{\textit{ab initio} {\gls{MD}}}
\newacronym{GCMC}{GCMC}{grand-canonical {\gls{MC}}}
\newacronym{HRMC}{HRMC}{hybrid {\gls{RMC}}}
\newacronym{XPS}{XPS}{X-ray photoelectron spectroscopy}
\newacronym{GW}{GW}{$GW$ theory}
\newacronym{ANN}{ANN}{artificial neural network}
\newacronym{KRR}{KRR}{kernel ridge regression}
\newacronym{MC}{MC}{Monte Carlo}
\newacronym{GPR}{GPR}{Gaussian process regression}
\newacronym[longplural={core-electron binding energies}]{CEBE}{CEBE}{core-electron binding energy}
\newacronym{ACO}{a-CO$_x$}{oxygen-rich amorphous carbon}
\newacronym{GO}{GO}{graphene oxide}
\newacronym{rGO}{rGO}{reduced graphene oxide}
\newacronym{RMC}{RMC}{reverse {\gls{MC}}}
\newacronym{XRD}{XRD}{X-ray diffraction}
\newacronym{vdW}{vdW}{van der Waals}
\newacronym{ASE}{ASE}{the Atomic Simulation Environment~\cite{larsen_2017}}
\newacronym{PAW}{PAW}{projector augmented-wave~\cite{bloechl_1994,kresse_1999}}
\newacronym{XANES}{XANES}{X-ray absorption near-edge spectroscopy}
\newacronym{SAXS}{SAXS}{small-angle X-ray scattering}
\newacronym{SI}{SI}{supporting information}

\begin{document}
\title{Experiment-driven atomistic materials modeling: A case study combining X-ray photoelectron spectroscopy and machine learning potentials to infer the structure of oxygen-rich amorphous carbon}

\author{Tigany Zarrouk}
    \email{tigany.zarrouk@aalto.fi}
    \affiliation{Department of Chemistry and Materials Science, Aalto University, 02150 Espoo, Finland}
\author{Rina Ibragimova}
    \affiliation{Department of Chemistry and Materials Science, Aalto University, 02150 Espoo, Finland}
\author{Albert P. Bart\'ok}
    \affiliation{Department of Physics, University of Warwick, Coventry CV4 7AL, United Kingdom}
    \affiliation{Warwick Centre for Predictive Modelling, School of Engineering, University of Warwick, Coventry CV4 7AL, United Kingdom}
\author{Miguel A. Caro}
    \email{mcaroba@gmail.com}
    \affiliation{Department of Chemistry and Materials Science, Aalto University, 02150 Espoo, Finland}

%\date{\today} % Leave empty to omit a date
\date{2 May 2024}

\begin{abstract}
\begin{center}
\textbf{Abstract}
\end{center}

An important yet challenging aspect of atomistic materials modeling is
reconciling experimental and computational results.
Conventional approaches involve generating numerous configurations through molecular dynamics or Monte Carlo
structure optimization and selecting the one with the closest match to experiment.
However, this inefficient process is not guaranteed to succeed. We introduce
a general method to combine atomistic machine learning (ML) with experimental observables
that produces atomistic structures compatible with experiment \textit{by design}. We use
this approach in combination with grand-canonical Monte Carlo within a modified Hamiltonian formalism,
to generate configurations that agree with experimental data and are chemically sound
(low in energy). We apply our approach to understand the atomistic structure of oxygenated amorphous
carbon (a-CO$_{x}$), an intriguing carbon-based material, to answer the question of how much oxygen
can be added to carbon before it fully decomposes into CO and CO$_2$.
Utilizing an ML-based X-ray photoelectron spectroscopy (XPS) model trained from $GW$
and density functional theory (DFT) data, in conjunction with an ML interatomic potential,
we identify a-CO$_{x}$ structures compliant with experimental XPS
predictions that are also energetically favorable with respect to DFT. Employing a network
analysis, we accurately deconvolve the XPS spectrum into motif contributions,
both revealing the inaccuracies inherent to experimental XPS interpretation
and granting us atomistic insight into the structure of a-CO$_{x}$.
This method generalizes to multiple experimental observables and allows for the elucidation of
the atomistic structure of materials directly from experimental data, thereby enabling
experiment-driven materials modeling with a degree of realism previously out of reach.
\end{abstract}

\maketitle

\section{Introduction}

One of the objectives of computational materials modeling is to infer the atomistic structure
of matter, both for the purpose of satisfying our curiosity about what matter ``looks like'' at the
atomic scale, and to obtain structure-property relations which might help in the design of useful
materials and molecules. However, real materials often have a complex structure, which might
set the time and/or length scales required for accurate simulation beyond the reach of \gls{DFT}.
Fortunately, in recent years data-driven approaches that ``learn'' the \gls{DFT} \gls{PES}
with high fidelity have been developed. These so-called \glspl{MLP} use \gls{ML}-based
techniques such as \glspl{ANN}~\cite{behler_2007,behler_2017} or
\gls{GPR}~\cite{bartok_2010,deringer_2021} (and the related \gls{KRR} method) to make
\gls{DFT}-quality predictions of energy
and forces for a small fraction of the CPU
cost~\cite{butler_2018,deringer_2019,zuo_2020,schutt_2020}.

Thanks to the flexibility of these \gls{ML} approaches, properties
other than energies and forces can also be predicted from the atomic
structure. Recent examples include electron
density~\cite{grisafi_2018,benmahmoud_2020} (or even wave
functions~\cite{gastegger_2020}), atomic
charges~\cite{bereau_2018,muhli_2021b,ko_2021}, molecular
dipoles~\cite{veit_2020,staacke_2022}, adsorption
energies~\cite{caro_2018c,jager_2018}, and spectroscopic
signatures~\cite{ghosh_2019}, X-ray spectroscopy in
particular~\cite{timoshenko_2017,aarva_2019,aarva_2019b,carbone_2020,golze_2022,
  kwon_2023,kotobi_2023,chen_2023}. Some of these predictions, specifically
spectroscopic calculations, are amenable to direct comparison with
experiment. This opens the door to designing new ways to sample
a material's configuration space, combining a computationally cheap and
accurate model of the \gls{PES} with improved recipes for how to
navigate it. This has the potential to provide a tight integration
between experimental data and
simulation~\cite{schwenker_2022,unruh_2022,kwon_2023b},
leading to a paradigm shift in materials modeling and
atomistic structure prediction. This new paradigm, that we call
``experiment-driven atomistic materials modeling'', is particularly
relevant to improve our understanding of complex and amorphous
materials.

In this paper, we focus on the integration of \textit{ab-initio}-accurate \glspl{MLP}
and \gls{ML}-based computational \gls{XPS} models to generate low-energy structures
which also have experimental \gls{XPS} agreement. First, we use classical atomistic
simulation methods based on \gls{MD} and \gls{MC} to sample the \gls{MLP} \gls{PES}
and show that the resultant \gls{XPS} predictions of final structures do not agree
well with experimental \gls{XPS} spectra. Then, we detail a method to combine these
models, by
performing an on-the-fly prediction and validation against the experimental
reference of the \gls{XPS} during structural optimization with \gls{MC}.
This ``generalized Hamiltonian'' or
``modified dynamics''
formalism results in atomistic structural models that produce
the same observables as the experiment \textit{by design}, while
ensuring the models remain structurally and energetically sound.

We present this method together with a case study concerning the elucidation of the atomistic
structure of \gls{ACO}. This material has potential applications such as in
memristors~\cite{santini_2015,bachmann_2017}, and by varying the degree of oxygenation
the tribological and electrochemical properties of oxygen-free
a-C~\cite{palomaki_2016,qasim_2019} could be tuned. Indeed, understanding a-C oxidation is key
to predict and potentially mitigate the long-term effects of friction on a-C coatings~\cite{marchon_1990}.
An important fundamental question pertaining to \gls{ACO} is also of interest:
how much oxygen can be added to carbon before it inevitably burns?
Although we focus here on \gls{XPS}
as our experimental target, because it is an experimental technique
widely used for the study of disordered materials, the methodology and
approach are general and extension to other experimental techniques
is straightforward. Our group is in active development with regards to these extensions.

\section{Interpretation of X-ray photoelectron spectroscopy}

\gls{XPS} is a ubiquitous experimental method
used to measure the distribution of \glspl{CEBE} in a
material. A specimen is irradiated with monochromatic X-rays which excite
core electrons such that they are photoemitted. The difference in the
measured kinetic energy of these electrons and the energy of the
incident X-ray gives a range of \glspl{CEBE}.
The usefulness of this method stems from the fact that the binding energy
of a core electron depends on its local environment, allowing for inference of the
structure of a material upon inspection of its \gls{XPS} spectrum.
\glspl{CEBE} of
reference environments---say an sp$^2$ carbon atom in graphite, or a
molecule where carbon is bonded to a certain functional group---can be
measured. Choosing a number of reference environments which one
expects in the material, and assuming that environments in the
material similar to those of the references give similar \glspl{CEBE},
one can approximately determine the composition of
an unknown structure by \emph{deconvolving} the measured
spectrum. Deconvolution splits a spectrum into peaks, centered
at the reference \glspl{CEBE}, which sum to give the
measured spectrum. The ratio of peak areas gives the relative proportion of
different environments. In this section we highlight some of the issues with
this approach and outline how simulations may help to overcome them.

\subsection{Reference peak energies are not transferable}
\label{sec:non_transferable_peak_energies}

We examine the validity of reference peak values commonly used in \gls{XPS} peak
fitting~\cite{moulder_1992} with an example from the literature. The issue
of peak assignment and practicalities of fitting \gls{XPS} curves has already been
discussed by others (see, e.g., Refs.~\cite{major_2020,gengenbach_2021,bagus_2023}).
We propose an alternative method for fitting in Sec.~\ref{sec:deconvolution_analysis} and show results in Sec.~\ref{sec:motifs_and_deconvolution}. 

\begin{figure}[t]
  \centering
  \includegraphics[width=\linewidth]{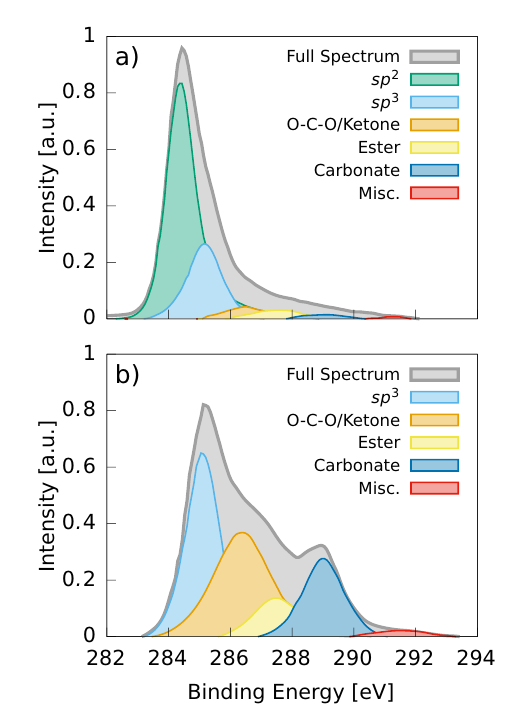}
  \caption{Example of experimental C1s XPS deconvolution of (a) an a-C structure
    (reported O:C ratio 0.1) and (b) a-CO$_x$ (reported O:C ratio 0.6) performed by
    Santini \emph{et al.}~\cite{santini_2015}, including the peak assignment proposed
    by the authors. We have subtracted the background
    by the method of Shirley~\cite{Shirley1972}. It can be shown within our simulation
    framework that the whole spectrum of (b) has been shifted towards larger C1s
    binding energies due to the presence of oxygen atoms, which shift the electrostatic
    potential in their immediate surroundings. This shift does not allow for
    sp$^2$ motifs to be detected by the deconvolution algorithm as the
    sp$^2$ reference is fixed.}
  \label{fig:experimental_xps}
\end{figure}

We are interested in elucidating the structure of oxygen-containing carbon materials, hence we inspect the work by Santini \textit{et al.}~\cite{santini_2015} on \gls{ACO}, 
which was also examined in our previous work~\cite{golze_2022}.
\fig{fig:experimental_xps}, replotted from the original results, shows their peak assignment
based on standard reference values for a-C and \gls{ACO}. The overall \gls{XPS}
C1s spectrum is reconstructed by by the sum of its contributions. These contributions are from sp$^2$ carbons, sp$^3$
carbons and carbons which are part of different O-containing functional groups. We will use \fig{fig:experimental_xps} to illustrate one of
the pathological issues associated to traditional \gls{XPS} peak fitting, namely that
standard references established for one class of materials are not transferable to other
(possibly more complex) materials.

\begin{figure}[t]
  \centering
  \includegraphics[width=\linewidth]{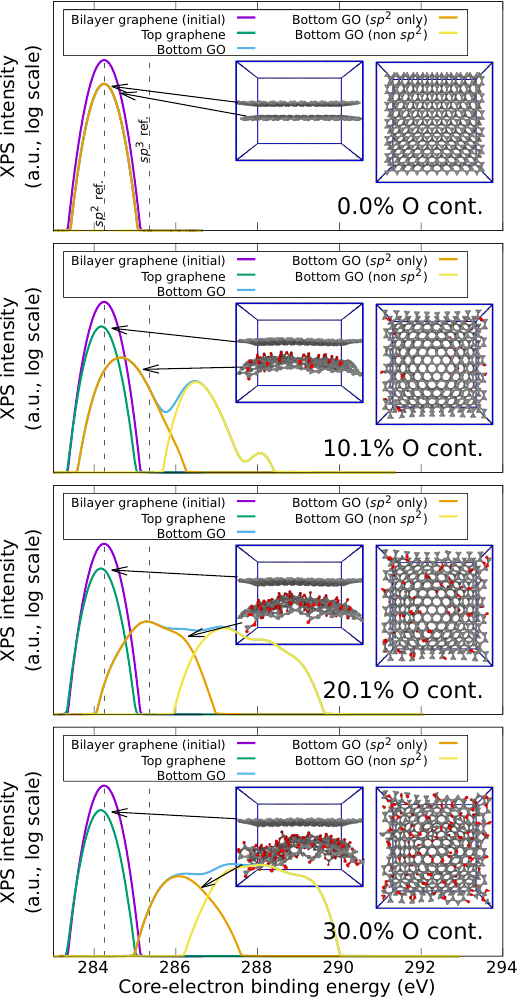}
  \caption{Simple computational experiment to illustrate how the sp$^2$ C1s peak
  shifts to higher \glspl{CEBE} as the amount of oxygen increases. The bottom layer in
  the initial bi-layer graphene structure (top panel) is progressively oxidized towards high O
  content (approx. 10~\% increase with each panel moving downwards in the graph) and its sp$^2$
  contribution to the \gls{XPS} spectrum shifts accordingly.}
  \label{02}
\end{figure}

Let us consider the following computational experiment, whose results are
depicted in \fig{02}. We start out with a
bi-layer graphene system, where the predicted \gls{XPS} spectrum can be trivially split into
the contribution of the upper layer and that of the bottom layer, since our \gls{ML} \gls{CEBE}
model~\cite{golze_2022} predicts individual \glspl{CEBE} which are unambiguously assigned
to individual atoms. Initially, the contributions to the overall spectrum from top and
bottom layers are identical. We proceed to add oxygen to the bottom layer in the form of
epoxide functional groups. These groups are only introduced with the oxygen pointing from the bottom
layer towards the middle of the bi-layer system so that possible
interactions with the top layer can be easily detected. To avoid generating unphysical structures, which could introduce
artifacts in the predicted \gls{XPS} spectrum, we proceed according to a \textit{heuristic}
approach inspired by Markov-chain \gls{GCMC}, where the potential energy of the system
is described with the \gls{MLP} introduced in Sec.~\ref{sec:mlp}. For context, we mark
the sp$^2$ reference peak position at $284.25$~eV and the sp$^3$ reference at $285.35$~eV,
i.e., the difference being 1.1~eV which is the commonly accepted splitting between sp$^2$ and sp$^3$ peaks in
a-C~\cite{nagareddy_2018}. We further split the \gls{XPS} spectra into the C1s contributions
from sp$^2$ motifs, defined as any given C atom bonded exclusively to three other C atoms,
and C atoms bonded to any number of neighbors one or more of which is an O atom (labeled as
``non sp$^2$'' on \fig{02}).
As more oxygen is incorporated into the resulting \gls{GO}
structure, we monitor the evolution of the
bottom and top \gls{XPS} spectra.

We can clearly see three effects: 1) the \glspl{CEBE} of the
top sp$^2$ C atoms largely remain at the reference value, this is what is expected according
to conventional \gls{XPS} peak fitting wisdom; 2) the C1s levels corresponding to C atoms
directly bonded to O atoms have significantly larger binding energies than pure carbon motifs,
also in accordance with conventional wisdom; 3) there is a very strong shift
of the sp$^2$ peak in the bottom layer towards higher
energies due to the presence of O, even though the
sp$^2$ C atoms remain bonded only to other C atoms. At high-enough O content ($\sim 20$~\%)
this shifted
sp$^2$ peak directly overlaps with the \textit{reference} sp$^3$ peak and, at even higher
O contents, it shifts to values significantly higher than the sp$^3$ reference. This
conflicts with the common practice of using immutable reference energies for \gls{XPS} peak
fitting. For the specific case of the experimental data in \fig{fig:experimental_xps}, we
can conclude that the peak attributed to sp$^3$ motifs in the a-CO$_x$ sample can be easily
explained by the presence of a strongly shifted sp$^2$ peak. In fact, as we will discuss
in Sec.~\ref{sec:structure_generation},
none of our attempts to generate computational a-CO$_x$ samples
led to a significant presence of sp$^3$ motifs in any of the structures, suggesting that
a-CO$_x$ is probably mostly sp$^3$-free. We note in passing that,
by virtue of the same argument, any possible sp$^3$ peak would anyway also be
significantly shifted towards higher energies with respect to the reference value. Values of this shifting will be shown in Sec.~\ref{sec:motifs_and_deconvolution}.

\subsection{Removing the ambiguity in deconvolution of spectra}

\gls{XPS} analysis is rife with deconvolution schemes, all giving different
peak proportions, muddying the deduction of material structure~\cite{major_2020,gengenbach_2021,bagus_2023}.
\gls{XPS} analysis of carbon
chars from Smith \emph{et al.}~\cite{smith_2016}
shows the pronounced effect of these deconvolution schemes on the
interpretation, predicting different amounts of carbon and oxygen
functional groups based on the deconvolution protocol used. Furthermore,
there is a lack of self consistency in many such approaches: the amount of oxygen
surmised from the deconvolution of the C1s spectra is not always consistent
with the actual amount of oxygen in the sample. The lack
of a universal scheme comes from multiple factors. Of note is that for
deconvolution an assumption is made: similar environments only
contribute to the \glspl{CEBE} within a narrow range
of the reference value. Given a motif structure in the material which is
similar to a reference (e.g., sp$^2$ carbon), which is now
influenced by other atomic species (e.g., environmental oxygen)
or defects in the local environment, there are \emph{shifts}
to the \glspl{CEBE} which \emph{cannot} be accounted for
by experimental deconvolution techniques, as the reference is fixed.
This is precisely the situation we exemplified in the previous section
with experimental \gls{ACO} data from the literature and our computational
\gls{GO} experiment.

Combining \gls{DFT} and experimental data, one can give estimates of the
\gls{CEBE} shifts for the \textit{molecular} references used in the
deconvolution~\cite{smith_2016}, however the issue remains: these
shifted molecular references will not have the same \glspl{CEBE} as
similar environments in the bulk material~\cite{aarva_2019}.
This is further complicated with regards to
the interpretation of amorphous materials, for which the references will have even less
similarity with the true environments present in the material. Thus, on the
one hand, a truly \textit{universal} library based on just a few reference
motifs cannot be made; on the other, traditional peak assignment leads to
highly arbitrary fits whenever too many reference peaks are used~\cite{gengenbach_2021}
(as famously put by John von Neumann~\cite{dyson_2004}, ``\textit{with four
parameters I can fit an elephant, and with five I can make him wiggle his trunk}'').

To obtain an unambiguous deconvolution of an \gls{XPS} spectrum and to avoid
the numerous pitfalls listed above, one can opt for another approach:
given a hypothetical atomic structure which matches the experimental sample,
one could use a computational model, ideally cheap to evaluate, to predict the
\glspl{CEBE} as a function of the local atomic environment, from which one
can readily obtain the \gls{XPS} spectrum. This spectrum is essentially a high-resolution
histogram of \glspl{CEBE} from a given atomic structure. The benefits of this
are multifold:
1) we know exactly each local environment contribution to the \gls{XPS} spectra;
2) each predicted \gls{CEBE} has the local environment shifts accounted for;
3) analysis of environment similarity in the structure allows for
 an \emph{unambiguous} deconvolution of the \gls{XPS} spectrum into motif
contributions.
In our previous work~\cite{golze_2022}, we showed that an \gls{ML} \gls{CEBE} model,
built on a careful combination of \gls{DFT} and $GW$ data, achieves the prescribed
requirements for quantitative accuracy and computational efficiency outlined above.
However, such a model still requires that the user provides an input atomic structure.
Therefore, the ``only'' problem that remains to be solved is:
how can one find a \emph{realistic} atomic structure whose predicted \gls{XPS} spectrum
matches the experimental one? Solving this problem requires a structure-generation
protocol that produces structural models which are \textit{simultaneously} low in
energy and reproduce the correct (experimental) \gls{XPS} spectrum.

\section{Matching atomic structure to experimental data}

The problem of matching a structure to
experimental data is within the realm of \gls{RMC}
techniques, first developed by McGreevy and Pusztai
\cite{McGreevy1988}, to find the structure of liquid argon
based on an
experimental pair-correlation function. These methods typically match
structure to experimental data only by moving atoms and evaluating a
Metropolis acceptance criterion. This criterion is based solely on a
measure of the agreement with
experiment for the given observable, thus without regard to whether
the structure is physical, i.e., low in energy for the given
thermodynamic conditions. To
circumvent this---without having to evaluate the total energy---the
addition of constraints is necessary. These range from user-defined,
material-specific, bonding constraints~\cite{biswas_2004} to adding
multiple sets of experimental data~\cite{limbu_2020}. However, these
models are typically under-constrained~\cite{opletal_2017}, and can
give unphysical results.
A natural way to determine the soundness of configurations during
simulation is to use the energy from an interatomic potential. This ensures the system remains in sensible regions of configuration space,
while matching experimental data. This strategy is followed in
\gls{HRMC} approaches~\cite{opletal_2013}, which have thus far
only been done by combining empirical potentials
and experimental observables, with the limitation that both must be amenable
to cheap computation with simple analytical expressions (e.g., pair-correlation functions and
\gls{XRD})~\cite{opletal_2013,opletal_2017,khadka_2020,farmahini_2013,bacilla_2022}.
A recent example of full \gls{HRMC} is its application
to amorphous calcium carbonate~\cite{nicholas_2024},
where the effective interactions could be well described with a relatively simple force
field and the inverted experimental data was based on \gls{XRD} analysis.
An alternative way to incorporate experimental diffraction data to structural optimization,
while retaining \gls{DFT} accuracy for the description of the \gls{PES}, is to combine
\gls{RMC} and \gls{DFT} heuristically, thus reducing the number of \textit{ab initio} steps
that need to be carried out, as done for a-Si and GeSeAg in Ref.~\cite{pandey_2016}.

The problems with empirical potentials have been
widely stated and will not be commented on here in detail.
This includes detailed comparisons of their
performance against the \gls{GAP} framework used here, even for the case
of simulating disordered carbon materials specifically~\cite{detomas_2019,caro_2020c,marchant_2023}.
In brief, these potentials cannot accurately reproduce the
\gls{PES} except for in those regions of configuration space for which they
were optimized. They are usually parametrized for specific materials,
and can fail spectacularly, or even catastrophically (``blow up'', in jargon),
when configurations are out of the scope of the fit. This is exacerbated when
performing \gls{MC} simulations,
which proceed by evaluating sequential energy differences. Hence, there will be
large errors in the acceptance criterion and consequently the accepted
configurations, which can lead to unrealistic structures. \glspl{MLP}, on the
other hand, can reproduce the \emph{ab initio} \gls{PES} accurately, given the inclusion of 
enough suitably chosen training configurations. A sufficiently
general potential would overcome the need for complex, user-defined,
material-specific constraints as the potential ``knows'' what configurations
are unphysical, which is reflected in the total energy. We note in passing
that poorly designed \glspl{MLP} are also prone to blowing up, and that this
is actually more common in \glspl{MLP} than in empirical potentials because of
the significantly more complex functional form of the potential.

We propose an alternative way to navigate configuration space
in order to generate structures which are consistent with experimental
\gls{XPS} data \emph{and} low in energy with regards to \emph{ab initio}
calculations, allowing for reliable and accurate \gls{XPS} deconvolution and
experimental structure determination. We use a machine-learned
\gls{XPS} model---which is informed by \gls{GW}---and a
CO \gls{GAP} to predict experimental structures of \gls{ACO}.
\gls{XPS} spectra predictions are matched to
experimental \gls{XPS} spectra by generalizing the \gls{HRMC}
approach to encompass \gls{GCMC}
simulations, using energies from the CO \gls{GAP} and \gls{XPS} spectral
dissimilarities as inputs. This synthesis of \gls{ML}
models and experimental data allows for structure
prediction at far larger scales than \gls{DFT}, all while having
\emph{ab initio}-level accuracy. Furthermore, this method allows one to dispense
with inaccurate deconvolution schemes which plague experimental \gls{XPS}
analysis, providing deeper structural insight into \gls{XPS} experiments.

\section{Materials and Methods}

\subsection{Machine learning potential}\label{sec:mlp}

To model the \gls{PES} of the carbon-oxygen system we rely on a
\gls{GAP} trained from data computed using \gls{PBE-DFT}. All the
\gls{DFT} calculations are done with the VASP
code~\cite{kresse_1996,kresse_1999}, with all the technical parameters
given in the \gls{SI}. The details of the \gls{GAP}
theoretical and methodological framework can be retrieved from the
literature~\cite{bartok_2015,deringer_2021}. Briefly, \gls{GAP} uses
\gls{GPR}/\gls{KRR} to learn and then predict the potential energy
landscape of a system of interacting atoms as a function of atomic
descriptors, usually two-body (2b), three-body (3b) and many-body (mb;
formally equivalent to an ensemble of 3b
descriptors)~\cite{musil_2021}. Our \gls{GAP} architecture for the CO
potential incorporates:
\begin{enumerate}
\item 2b descriptors with 4.5~\AA{} cutoff for the C--C, C--O and O--O
  interactions;
\item 3b descriptors with 2~\AA{} cutoff for the six possible
  permutations of C and O triplets, where a central atom is singled
  out and the descriptor is invariant with respect to permutations of
  same-species atoms for the other two (CCC, CCO, COO, OCC, OCO and
  OOO, with the first symbol indicating the central atom);
\item mb soap\_turbo descriptors~\cite{caro_2019}, a modification of
  the \gls{SOAP} descriptor, with 4.5~\AA{} cutoff. One \gls{SOAP}
  descriptor is added for each species, with both species
  visible---i.e., the carbon-centered \gls{SOAP} descriptor is
  sensitive to both C and O neighbors, and so is the oxygen-centered
  \gls{SOAP} descriptor;
\item Tabulated ``core'' potentials for the C--C, C--O and O--O
  interactions, explicitly describing the highly repulsive regime when
  two atoms are down to 0.1~\AA{} from each other. The explicit inclusion
  of this term improves the stability and accuracy of the \gls{GAP}
  fit significantly;
\item Optionally, tabulated long-range pair potentials to describe, in
a limited way, \gls{vdW} (or ``dispersion'') interactions. Currently,
long-range C--C, C--O and O--O dispersion energies use a fixed parametrization.
This functionality will be extended to Hirshfeld volume prediction
so that environment-dependent \gls{vdW} correction schemes can be used,
as discussed in Ref.~\cite{muhli_2021b}.
\end{enumerate}

We refer to this \gls{MLP} as the CO-GAP. 
The CO-GAP furnishes us with \textit{local energy} predictions for each atom $i$ which are defined as 
\begin{align}
  \label{eq:local_energy}
  E^{i}_{\text{GAP}} ( &\{r\}, \{q^{\text{2b}}_i\}, \{\textbf{q}^{\text{3b}}_i\}, \mathbf{q}^{\text{mb}}_i) = 
   E^{\text{2b}}(\{q^{\text{2b}}_i\}) + E^{\text{3b}}(\{\textbf{q}^{\text{3b}}_i\}) \nonumber\\
  & + E^{\text{mb}}(\mathbf{q}^{\text{mb}}_i) 
  + \sum_{r < r_{\rm cut}^{\rm core}} E^{\text{core}}(r) + \sum_{r < r_{\rm cut}^{\rm disp}} E^{\text{disp}}(r) \nonumber
\end{align}
where $\{r\}$ is a set of atomic pair distances with respect to atom $i$, $E^{\text{core}}(r)$ and $E^{\text{disp}}(r)$ are energies from the core potential and dispersion interactions, respectively, $\{q^{\text{2b}}_i\}$ and $\{\textbf{q}^{\text{3b}}_i\}$ are sets of 2b and 3b descriptors pertaining to each pair or triplet of species, respectively, and $\mathbf{q}^{\text{mb}}_i$ is a soap\_turbo descriptor. The total energy of a configuration is simply given by the sum of local energies: 
\begin{equation}
    E_{\text{GAP}} = \sum_i E^{i}_{\text{GAP}} ( \{r\}, \{q^{\text{2b}}_i\}, \{\textbf{q}^{\text{3b}}_i\}, \mathbf{q}^{\text{mb}}_i).\nonumber
\end{equation}  

This CO-GAP is freely available from
the Zenodo repository~\cite{caro_2024}. Future
(improved) versions of the CO-GAP will also be added to the same repository, which
will retain the whole version history to ensure reproducibility of published
results. The CO-GAP can be used to predict energy and forces with
TurboGAP~\cite{ref_turbogap}, QUIP~\cite{csanyi_2007,ref_quip} and
LAMMPS~\cite{plimpton_1995,ref_lammps} via its QUIP interface. The modified dynamics
with on-the-fly \gls{XPS} prediction and inclusion of experimental observables
discussed in Secs.~\ref{sec:combining} and \ref{sec:modified_gcmc} is only currently
available in TurboGAP.

\subsection{Iterative database generation}

\begin{figure*}
  \centering \includegraphics[width=\textwidth]{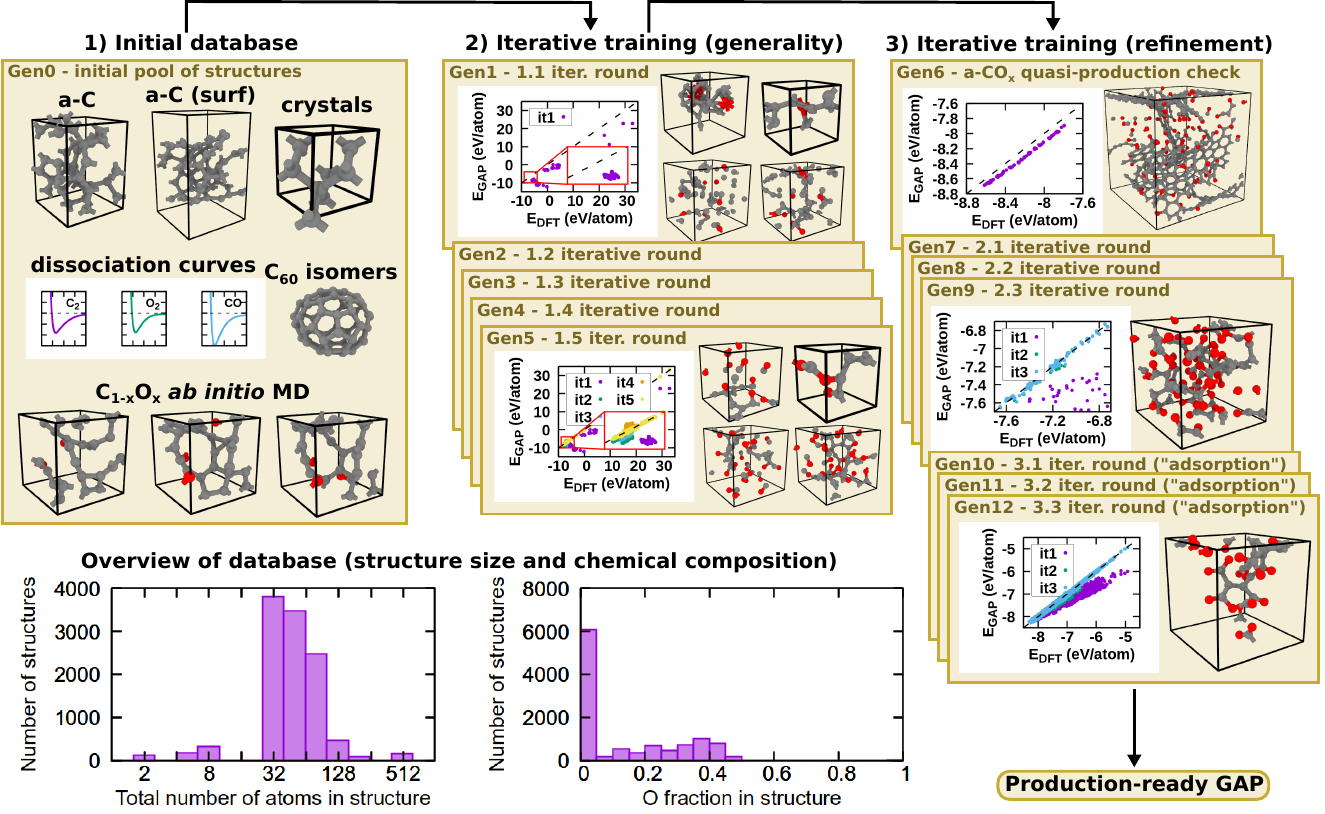}
  \caption{Iterative training workflow followed while developing the CO-GAP. The
  initial database, which is added for stability and generality, is
  incrementally improved and refined by
  adding more purpose-specific and high-O-content structures. The two panels in
  the bottom-left corner show an overview of the composition of the database,
  in terms of the size (number of atoms) of the periodic simulation boxes and
  their stoichiometry.}
  \label{01}
\end{figure*}

The initial training database is constructed by using the a-C database of
Deringer and Cs\'anyi~\cite{deringer_2017}, recomputed at the
\gls{PBE-DFT} level. We extend it by adding all
possible C$_{60}$ isomers~\cite{sure_2017}, the dissociation curves of
CO and O$_2$, and iteratively generated a-C:O structures of different
sizes and compositions. The iterative training procedure is as
follows. Initially, before the \gls{GAP} has ``seen'' any O-containing
carbon structures (beyond the CO dissociation curve), we run three
independent \gls{AIMD} trajectories with different mass densities.
Initial $3 \times 3 \times 3$ simple cubic structures, containing 24 C
atoms and 3 O atoms randomly substituted in the lattice, are quickly
quenched from 4000~K down to 300~K over 1~ps. This provides a
computationally affordable glimpse at the configuration space spanning
from l-C:O down to a-C:O. Ten snapshots from each run are chosen to be
included in the training database. A first version of the CO-GAP
is fitted from this database, and used instead of \gls{DFT} to cheaply
generate more training data as in the first iteration, but this time
sampling 9 different starting configurations varying from 8 up to 64
atoms at a similar composition ($\sim 10$~\% O content). These
structures are quenched from 5000~K down to 300~K over a significantly
longer 20~ps simulation, collecting snapshots every 2~ps, to avoid
highly correlated configurations. These single-point snapshots are
used to run \gls{PBE-DFT} calculations and added to the growing
training database. The same procedure is repeated several times, while
slowly increasing the maximum O content, until the \gls{GAP} generates
low-temperature structures of arbitrary composition whose predicted
energies are close enough to the \gls{DFT} values.

This first iteratively trained \gls{GAP} is used in a series of
quasi-production runs, with 210 512-atom simple-cubic structures at
different densities (1.5 to 2.5~\unit{\gram\per\cubic\centi\metre}) and compositions (2.5 to
60~\% O contents), to generate \gls{ACO} structures using a slower
quench process~\cite{caro_2023} from 3500~K down to 300~K over
100~ps. The final configurations range in structure from a-C:O to
a-CO$_x$ and even ``burnt'' systems, with lots of CO and CO$_2$
molecules spontaneously forming at very-high O content. They are
recomputed with \gls{PBE-DFT} and also added to the training database.
The resulting \gls{GAP} still undergoes a few rounds of iterative
training for high-O-containing structures with a CO/CO$_2$ removal
step at the end (before the \gls{DFT} calculation). Finally, another
few rounds of iterative training are carried out to include
``adsorption'' configurations, i.e., \gls{ACO} structures derived by
adding O atoms to a pre-existing graphitic a-C structure on the bridge
or top positions, corresponding to the classical ether/epoxide and
ketone organic chemistry groups, respectively. In these adsorption
iterative rounds, one O atom is placed at the adsorption site, followed
by a relaxation and removal of any possible CO and CO$_2$ formed
molecules, before placing the next O atom and repeating the procedure.

One of the advantages of this iterative training procedure is that the
\gls{DFT} calculations used to train the next generation of the
\gls{MLP} can be used to compute an unbiased error estimate for the
previous generation of the \gls{MLP}. The whole workflow to train the
CO-GAP is depicted in \fig{01}, including example structures and
scatter plots for selected \gls{GAP} versions, where \gls{GAP}
predictions are tested for the structures to be added to the
\textit{next} generation of the training database. Overall, our
training database was incrementally enhanced from generation zero
(``Gen0'') until generation 12 (``Gen12''). Going from one generation
to the next involves training a new \gls{GAP}. Thus, the last
``production-ready'' \gls{GAP}, trained from the Gen12 training
database, is the 13th \gls{GAP} overall that was trained as part of
the iterative process.
The histograms on the bottom-left corner of the figure show the overall
composition of the \textit{final} training database, in terms of total
number of atoms per structure and structure stoichiometry. We note
that most of the training configurations have between 32 and 128
atoms, with some very small (2, 4 and 8 atoms) and very large (512
atoms) structures. We also note that a large fraction of structures
contain only carbon atoms, taken from the initial training
database. This is required because of the large diversity of carbon
structures that can be realistically found in nature (in particular in
a-C). The relevant size of configuration space for O-containing
structures is comparatively small, because there are not so many ways
in which O atoms can bind to C atoms to form stable arrangements.

All the fits were carried out with the \texttt{gap\_fit}
program~\cite{klawohn_2023,klawohn_2023b}, part of the QUIP software
package~\cite{csanyi_2007,ref_quip}. Atomic structure generation, manipulation, \gls{MD}
simulations, etc., were done with \gls{ASE}, different in-house
codes and the TurboGAP program~\cite{ref_turbogap}.

\subsection{Computational X-ray spectroscopy}

To predict \gls{XPS} spectra from \gls{ACO}, we use the
\gls{SOAP}-based \gls{XPS} model that we have previously developed~\cite{golze_2022},
retrained with \texttt{gap\_fit}'s ``local property'' fitting functionality~\cite{klawohn_2023}.
This model predicts \glspl{CEBE} based on the local atomic environment of an
atom $i$, which is characterized by a \gls{SOAP} descriptor $\mathbf{q}_i$:
\begin{equation}
  \label{eq:soap_xps}
  \bar{E}^\text{sp} (\mathbf{q}_i) = E_0^\text{sp} + \delta^2 \sum_s \alpha_s^\text{sp} k( \mathbf{q}_i, \mathbf{q}_s),
\end{equation}
where $\bar{E}^\text{sp} (\mathbf{q}_i)$ is the \gls{CEBE} prediction,
$\delta^2$ provides the energy scale,  $\{\alpha_s\}$ is the set of fitting coefficients,
and $E_0$ is an offset energy, ideally chosen close to the training set average.
The sp superscript (for ``species'') indicates that different models are trained for
different species, in our case we can train one model for C1s \glspl{CEBE} and
another one for O1s \glspl{CEBE}; while this paper focuses on the former, the model
architecture does not change, and our previous paper discusses both~\cite{golze_2022}.
In \eq{eq:soap_xps}, $\{ \mathbf{q}_s \}$ is the set of \gls{SOAP} descriptors in the
sparse set, $s \in \{1,\ldots,N_{\rm s} \}$, where $N_{\rm s}$ is the number of
sparse configurations (see Ref.~\cite{deringer_2021} for a discussion of sparse \gls{GPR}
in the context of atomistic \gls{ML}). $k(\mathbf{q}_i, \mathbf{q}_s)$ is the kernel
function, which gives the similarity between the atomic environments of atoms $i$ and $s$,
ranging between 0 and 1 and given by
\begin{equation}
  \label{eq:kernel}
  k( \mathbf{q}_i, \mathbf{q}_s) = \left( \mathbf{q}_i \cdot \mathbf{q}_s \right)^{\zeta},
\end{equation}
where $\zeta = 2$ in our case. In the following, we use $\bar{E}_i \equiv \bar{E}^\text{sp} (\mathbf{q}_i)$
for shorthand notation.

The training set was composed of \glspl{CEBE} which
were calculated, using a combination of \gls{DFT} and $GW$ reference data,
from bulk material as well as surfaces. This model accounts for the fact that in
solid-state samples the experimental reference energy is given by the Fermi level
of the material (as opposed to the vacuum level used for molecular samples).
The training database consisted of a-C, a-C:O, functionalized
a-C, \gls{ACO}, graphene, and \gls{rGO}.
For further details, see Ref.~\cite{golze_2022}.

To account for thermal and instrumental broadening, the predicted \glspl{CEBE} were
broadened by a Gaussian of width $\sigma = 0.4$~eV, and the resulting data was normalized
over the domain of prediction, giving the final \gls{XPS} spectrum from the contributions
of all atoms contained in a computational structural model $S$:
\begin{equation}
  \label{eq:spectrum_contribution}
 g_{\rm {pred}}(E; \{\mathbf{q}_i\}) = \frac{1}{M} \sum_{i \in S}
 \exp\left( -\frac{\left(E - \bar{E}_{i}\right)^2}{2\sigma^2}\right),
\end{equation}
where $E$ is an arbitrary energy along the domain of the spectrum and
$M$ is a normalization factor to make
$\int \,\text{d}E \, \left( g_{\rm {pred}}(E) \right)^2 = 1$.
While the experimental peak fitting procedure uses peak widths
as a way to account for instrumental broadening, thermal broadening \textit{and}
structural disorder, our procedure incorporates disorder directly by generating the
distribution of atomic environments explicitly. Therefore, our smearing parameter
should only account for instrumental and thermal broadening, which should be fixed
for given experimental conditions: i.e., given an experimental apparatus and temperature,
the induced broadening that we should consider to mimic the convolution of both effects
is fixed. Our methodology implicitly incorporates the broadening in the \gls{XPS} spectrum
necessary to account for disorder by explicitly generating the necessary distribution
of structural motifs. Thus, within our method,
the $\sigma$
value is chosen based on the broadening that can be found in typical experimental
setups, where
synchrotron-based \gls{XPS} spectra will have better resolution that lab-based spectra
(see Ref.~\cite{aarva_2019b} for a discussion of $\sigma$ in the context of comparing
simulation to experiment).
We expect that the effect of too small/large
$\sigma$ values would be for the structure-generation protocol to counteract it by
inducing a broadening/narrowing of the spectral features by spuriously
exacerbating the degree of atomic motif disorder/order in the structure.
For these reasons, $\sigma$
is \textit{not} an optimizable free parameter of the model but must be chosen on physically
motivated grounds.
We further discuss the choice of $\sigma$ and show the effect of varying this parameter in the
\gls{SI}.

\subsection{Combining experiment and simulation for realistic structure determination}\label{sec:combining}

We can compare the normalized \gls{XPS} spectrum
$g_{\rm {pred}}(E; \{\mathbf{q}_i\})$ to the normalized experimental spectrum $g_{\rm {exp}}(E)$,
since \textit{absolute} \gls{XPS} intensities are arbitrary. We can define a ``dissimilarity'' metric, $\mathcal{D}$,
that varies from 0 to 1, which we take to be a sum of squared differences:

\begin{align}\label{eq:dissimilarity}
  \mathcal{D} \left( \left\{\mathbf{q}_i\right\}
  \right) & = \frac{1}{2} \int \text{d}E \left(  g_{\rm {pred}}(E; \{\mathbf{q}_i\}) -  g_{\rm {exp}}(E)  \right)^2
  \nonumber \\
  & = 1 - \int \text{d}E \, g_{\rm {pred}}(E; \{\mathbf{q}_i\})
  \, g_{\rm {exp}}(E),
\end{align}
where the second equality is valid because of the normalization
of the spectra.

Defining an energy scale as $\gamma$, we can create the pseudoenergy
$E_{\rm spectra}(\{\mathbf{q}_i\}) \equiv \gamma\mathcal{D}\left( \left\{  \mathbf{q}_i  \right\}
\right)$, which allows
for construction of the modified total energy of the system
\begin{equation}\label{eq:etilde}
  \tilde{E} := E_{\rm pot} + E_{\rm spectra},
\end{equation}
where $E_\text{pot} \equiv E_\text{GAP}$ (in our case) is the potential energy of
the system. Although the focus in this paper is on \gls{XPS}, we could more generally
define the modified energy from an arbitrary
linear combination of contributions, each based on an experimental observable, as
$\tilde{E} := E_{\rm pot} + \sum_j \gamma_j E_j$.

$E_{\rm spectra}$ acts as a penalty term, increasing the
energy with increasing spectral dissimilarity. From a \gls{RMC} perspective,
where the focus is on matching experimental data
regardless of atomic total energies, the $E_{\rm pot}$
term acts as an energy-based constraint on the atomic positions: high-energy
atomic configurations are penalized. By optimizing $\tilde{E}$
with an appropriate $\gamma$, we perform a multi-objective optimization
with respect to the atomic positions, favoring the generation of
atomic configurations which \textit{simultaneously} match the experimental \gls{XPS} spectrum and are low in energy.
An additional benefit of this multi-objective optimization is that
the \gls{CEBE} model only needs to be quantitatively accurate for low-energy structures,
as extrapolation errors for high-energy structures do not impact the optimization because
the potential energy constraint discourages these structures from being accepted. This is
illustrated in Fig.~S5 of the \gls{SI}.

For computational efficiency, we use the same structural mb descriptors $\{\textbf{q}_i\}$
for the \gls{CEBE} model as we do for the CO-GAP. This speeds up the simulation:
the $E_{\rm pot}$ term is calculated first, which requires a descriptor calculation for
each atom in the structure. This step has the highest computational cost incurred during
simulation. As the relevant descriptors have already been calculated,
the only necessary calculation for prediction of the \gls{XPS} spectra is the evaluation
of the kernel function between these descriptors and those of the sparse set corresponding
to the \glspl{CEBE}, \eq{eq:soap_xps}.

One of the purposes of \gls{XPS} analyses in multispecies materials is to determine the
elemental composition of the samples. Therefore, being able to reliably tune the relative number of atoms of different species in our simulations is essential.
For instance, the \gls{XPS} analyses of \gls{ACO} by Santini \emph{et al.}~\cite{santini_2015}
show that the O:C ratio varies significantly with O$_2$ partial pressure. For comparison,
we must generate representative structures at a range of oxygen concentrations. A natural
way to generate these structures is to start from pure a-C and perform \gls{GCMC} simulations
with different chemical potentials with our modified energy $\tilde{E}$, as discussed next.

\subsection{Modified grand-canonical Monte Carlo}\label{sec:modified_gcmc}

In a \gls{GCMC} simulation, a system of interest is at fixed volume
$V$, allowed to thermalize by contact with a heat bath at temperature
$T$ and exchange particles with an infinite reservoir at a chemical
potential $\mu$, forming a constant ($\mu,V,T$) ensemble. We perform
\gls{GCMC} using a Markov chain: starting from an initial pure a-C
structure, we generate trial configurations by either randomly
displacing a particular atom, or inserting/removing oxygen into/from a
random position, respectively. These trial configurations are either
accepted or rejected using the standard acceptance criteria~\cite{frenkel_2002}
for particle displacement/insertion/removal:
\begin{align}
\pushleft{\mathrm{acc}( \mathrm{move}) =}
\nonumber \\
\pushright{\mathrm{min}\biggl[1,
\exp\left\{ -\beta ( \tilde{E}(\mathrm{trial}) - \tilde{E}(\mathrm{current}) ) \right\} \biggr],}
\label{eq:gcmc_move}
\\
\pushleft{\mathrm{acc}(N \rightarrow N+1) =}
\nonumber \\
\pushright{\mathrm{min}\biggl[ 1,
\frac{V}{\lambda^3 (N+1)} \exp\left\{ - \beta ( \tilde{E}(N+1) - \tilde{E}(N) - \mu ) \right\} \biggr],}
\label{eq:gcmc_add}
\\
\pushleft{\mathrm{acc}(N \rightarrow N-1) =}
\nonumber \\
\pushright{\mathrm{min}\biggl[1,
\frac{\lambda^3 N}{V} \exp\left\{ -\beta ( \tilde{E}(N-1) - \tilde{E}(N) + \mu ) \right\} \biggr],}
\label{eq:gcmc_remove}
\end{align}
where $\lambda$ is the thermal de-Broglie wavelength, given by
$\lambda = \sqrt{\frac{2\pi \hbar^2}{m k_\text{B} T}}$,
and $\beta = 1 / k_\text{B}T$. A trial configuration
is accepted if the corresponding acceptance criterion is greater than
a random number $r \in [0,1]$. We then repeat the procedure with the
last accepted configuration until the maximum number of iterations
has been reached.

Thus, the steps for performing \gls{GCMC} simulations to optimize the
\gls{XPS} spectrum are as follows:
\begin{enumerate}
\item Generate trial configuration from a randomly chosen \gls{MC}
step type (insertion, removal, move);
\item Evaluate $\tilde{E}$, from \eq{eq:etilde};
\item Evaluate the corresponding acceptance criterion, one of Eqs.~\eqref{eq:gcmc_move},
\eqref{eq:gcmc_add} and \eqref{eq:gcmc_remove}, and compare to $r$;
\item Repeat until the target number of \gls{MC} iterations has been reached.
\end{enumerate}

\subsection{Deconvolution of {a-CO$_x$} spectra}\label{sec:deconvolution_analysis}

We deconvolved C1s XPS spectra by considering the resultant structures of
\gls{GCMC} simulations as undirected graphs: atoms were nodes
differentiated by species, and bonds were edges. This was achieved
with the NetworkX package in
Python~\cite{SciPyProceedings_11} in conjunction with \gls{ASE}~\cite{larsen_2017}. Bonds were defined by overlaps of atomic spheres, where an atomic radius of 0.912~\AA{} was used for C and a radius of 0.792~\AA{} for O---this was done using the in-built \gls{ASE} functions \texttt{NeighborList} and \texttt{natural\_cutoffs} functions with \texttt{mult = 1.2}. For each carbon, a graph
was made which contains its first nearest neighbors and, if an oxygen was found, the neighbors of the oxygen. A reference graph database was created from the connectivity of standard motifs (e.g., sp, sp$^2$,
ether, etc.). Subgraph isomorphisms were sought between the reference
graphs and each carbon subgraph in the structure in a hierarchical fashion, from the
most complex to the least (in the order: carbonate, peroxide, ester,
epoxide, ether, ketone, CO$_2$, CO, sp$^3$, sp$^2$ and sp), with the most complex subgraph isomorphism being chosen. This was
sufficient to associate each carbon with a motif. C1s
\glspl{CEBE} were grouped according to these motifs,
allowing for the delineation of each motif contribution to the \gls{XPS}
spectrum. 

\section{Structural model generation}\label{sec:structure_generation}

In this section we explore different approaches to generating \gls{ACO} and assess their
ability to yield results in agreement with experiment. In particular, the challenge in
\gls{ACO} atomistic structure prediction is to generate metastable oxygenated a-C structures
with very high O content. Although the most realistic strategy would be to
carry out simulated deposition as has been done for a-C in the
past~\cite{marks_2005,caro_2018,caro_2020c,caro_2023}, trying to mimic the
experimental growth process, this is also the most sophisticated and time consuming method,
in terms of both human and computer time required. Therefore, we explore alternative, cost
effective, strategies and compare them to the modified \gls{GCMC} approach developed here.

\subsection{Melt-quench generation}\label{sec:melt_quench}

\begin{figure}
  \centering
  \includegraphics[width=1\columnwidth]{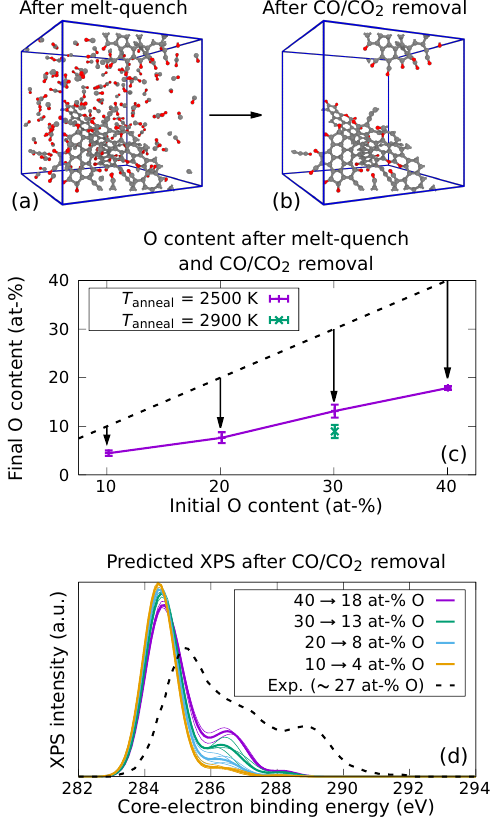}
  \caption{Example CO melt-quench sample after (a) high-temperature annealing and
  quenching and (b) subsequent removal of molecular species. (c) Final O content in the
  samples after CO/CO$_2$ removal as a function of the original O content. (d) Predicted
  XPS corresponding to the melt-quench samples after molecular species removal, compared to
  the target experimental spectrum, where we have indicated the O content inferred with our
  modified GCMC approach in Sec.~\ref{sec:gcmc_results}.}
  \label{fig:melt-quench}
\end{figure}

For a-C, melt-quench \gls{MD} simulations are arguably the most straightforward type of atomistic
structure generation procedure, after a simple geometry
optimization~\cite{marks_1996,marks_2005,caro_2014,laurila_2017,caro_2023}.
Melt-quench simulations of a-C start out from a high-temperature liquid sample (``melt'')
and rapidly cool it down (``quench'') to trap the atoms into metastable configurations. There
is an intermediate high-temperature annealing step, where the system is allowed to equilibrate
and find more energetically favorable configurations. Depending on the duration of the
annealing step and the quenching rate, disordered carbons with different degrees of graphitization
are obtained~\cite{detomas_2016,deringer_2017,detomas_2019,jana_2019,wang_2022,caro_2023}.
In addition to the choice of a temperature profile with an \gls{MD} thermostat,
whether the density of the system is fixed or an external pressure is applied via
an \gls{MD} barostat will also affect the properties of the resulting material.

\begin{figure*}[t]
  \centering \includegraphics[width=1\textwidth]{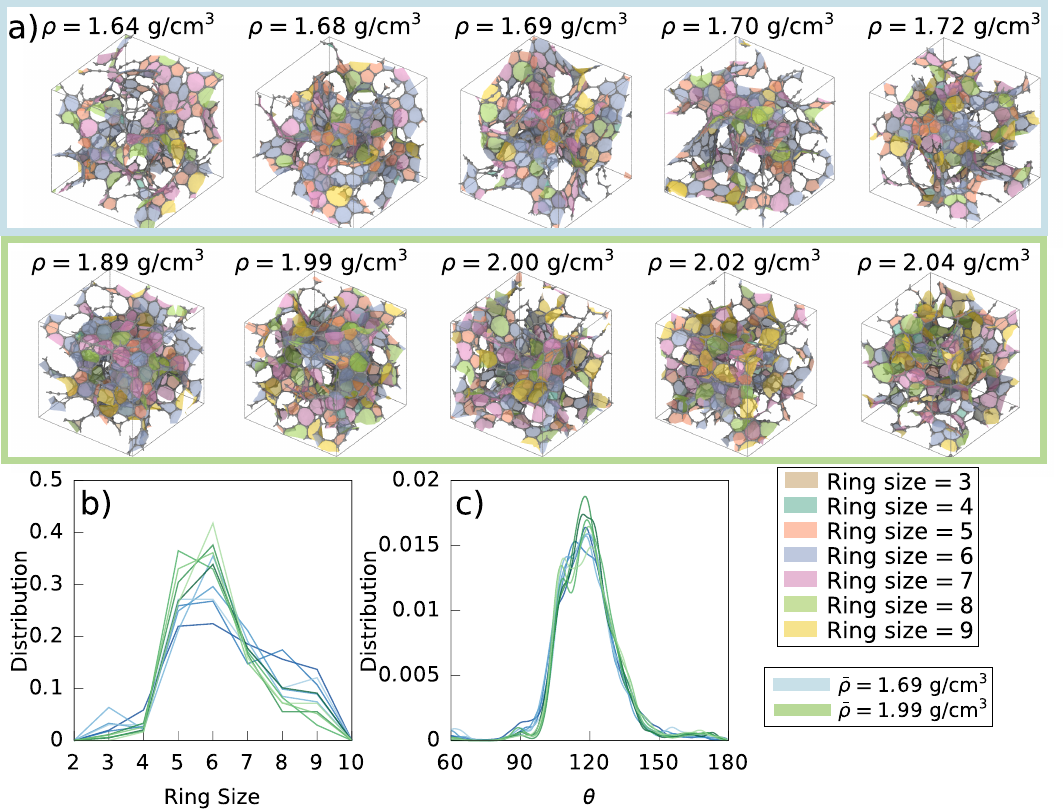}
  \caption{The two groups of initial structures used in the simulations, which
    have distinct densities. a) Structures with surfaces colored
    according to the number of members in the carbon ring. The ring search was capped at 9-membered rings. b) Ring-size distributions.
    c) Bond-angle distributions. Saturation of color enumerates the structures in the group, the lightest being first and the darkest being last. }
  \label{fig:structures_bond_angle}
\end{figure*}

Given the past success and popularity of melt-quench simulations for a-C structure generation,
a naive attempt at generating computational a-C:O and \gls{ACO} samples using this method
appears as an obvious
choice. Therefore, we carried out melt-quench simulations of a-C:O formation with our CO-GAP.
We start with randomly initialized systems with fixed C:O ratios (10, 20, 30 and 40 at-\% O
contents) and apply the following temperature profile, while applying a 1~bar barostat:
10~ps of liquid thermalization at 5000~K, 100~ps of high-temperature annealing at 2500~K,
and 10~ps of low-temperature annealing at 300~K. At each C:O ratio 3-4 samples are simulated
to assess the effect of random initialization on the results. Invariably, the structures that
we obtained consisted of graphitic flakes with O passivation along the edges and a large degree
of CO and CO$_2$ formation. An example structure is shown in \fig{fig:melt-quench} before (a)
and after (b) the CO and CO$_2$ molecules were removed. The trends show that most of the
O atoms originally present in the simulation box are being used to generate the
thermodynamically more stable molecules, rather than being trapped into the metastable
amorphous motifs (\fig{fig:melt-quench} (c)). We repeated a series of simulations at a
higher annealing temperature (2900~K), clearly showing an even more pronounced formation
of molecular species.

As expected, the samples yield, after removing the molecular species,
\gls{XPS} signatures that are very different from Santini's
experimental sample which we aim to reproduce (\fig{fig:melt-quench} (d)), for which we
estimate in Sec.~\ref{sec:gcmc_results}, with our modified \gls{GCMC} simulations, an O content
$\sim 27$~\%. Thus, melt-quench simulations of CO mixtures
seem to favor the formation of the thermodynamically stable molecular products and produce
low-O-content a-C:O structural models, below $\sim 20$~at-\% O.
The inability of high-temperature annealing protocols to produce
experiment-compliant structures draws parallels with how simulations of undoped tetrahedral a-C
systematically failed to generate high-sp$^3$ samples matching experiment until high-accuracy
explicit deposition simulations became feasible with MLPs~\cite{caro_2018,caro_2020c,caro_2023}.
In the case of a-C:O and \gls{ACO}, melt-quench protocols further exacerbate this issue because
the favoring of high-entropy thermodynamically stable products in the C-O system leads to
molecular species formation, far away from the target solid-state samples, whereas for pure
carbon it simply favors solid-state samples with higher graphitic content (i.e., lower
sp$^3$/sp$^2$ ratios than in experiment), as the thermodynamically stable product is graphite.
Therefore, melt-quench
simulations are not a good surrogate for the (non-equilibrium) deposition mechanism
that takes place during the experimental \gls{ACO} synthesis. Hence we developed the
modified \gls{GCMC} scheme, which is detailed next.

\begin{figure*}[t]
\includegraphics[width=0.8\linewidth]{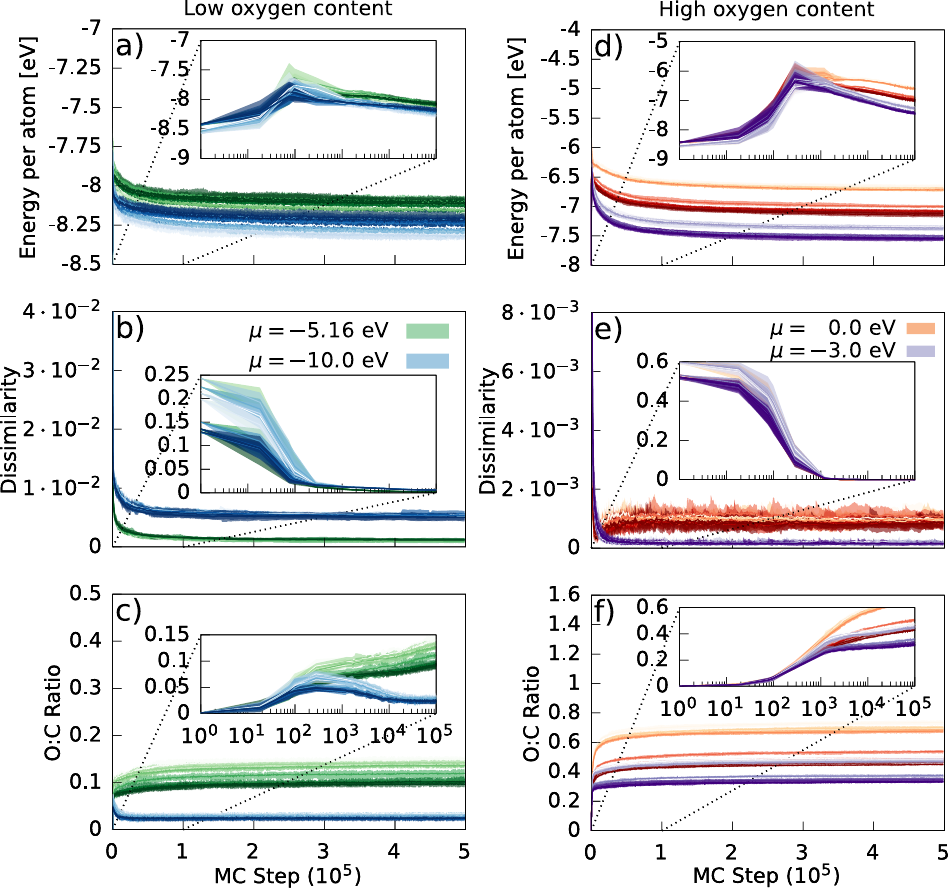}
\caption{Modified GCMC runs for different chemical potentials to
generate a-CO$_x$ structures, with \gls{XPS} optimization for the
low- (left-hand-side panels) and high-oxygen-content (right-hand-side panels)
\gls{XPS} spectra. Each line represents a different
structure. Solid lines are the averages over the runs at that
particular timestep, with shaded regions denoting the extrema of
the data. a)/d) Energy per atom. b)/e) Dissimilarity. c)/f) O:C
ratio. Inset are logscale plots, showing the initial equilibration
of the system.}
\label{fig:gcmc_runs_low_oxygen}
\label{fig:gcmc_runs}
\end{figure*}

\subsection{Modified and unmodified GCMC}\label{sec:gcmc_results}

\subsubsection{Choice of initial structures}

Ten 512-atom a-C structures were used for the initial configurations of
the modified \gls{GCMC} simulations, as seen in \fig{fig:structures_bond_angle} a).
These were generated from a melt-quench procedure, similar to Ref.~\cite{wang_2022}.
The structures can be grouped into two
distinct average densities, $\bar{\rho} \approx 1.69$~\unit{\gram\per\cubic\centi\metre} and
$\bar{\rho} \approx 1.99$~\unit{\gram\per\cubic\centi\metre}. 
The structures of lower density have
a slightly more diverse ring structure, as seen in
\fig{fig:structures_bond_angle} b). As such, their bond-angle
distribution differs slightly from those of higher density, in
\fig{fig:structures_bond_angle} c).

Ten separate Monte-Carlo \gls{XPS} optimizations were run for $500,000$ steps
for each structure at a given chemical potential
$\mu \in \{-10, -5.16, -3, 0\}$~eV, with a temperature of
300~K. The chemical potential of $-5.16 $~eV can be related to a
reference of half the \gls{DFT} binding energy of diatomic oxygen at
1~atm of pressure at 300~K, and is taken from Samanta \emph{et al.}
\cite{samanta_2010}. Two C1s \gls{XPS} spectra of a-CO$_x$ were used for
fitting, both taken from Santini \emph{et al.} \cite{santini_2015}:
for the chemical potentials of $\mu \in \{-10, -5.16\}$~eV, the XPS
spectrum of low oxygen content---O:C ratio $\approx 0.1$, according to
Ref.~\cite{santini_2015}---was used with $\gamma = N_{\rm C}\gamma_{\rm C}^{\text{low-O}}$ where $ \gamma_{\rm C}^{\text{low-O}}= 39.0$~eV per carbon atom and $N_{\rm C}$ was the number of carbon atoms;
for the chemical potentials of $\mu \in \{-3, 0\}$~eV, the \gls{XPS}
spectrum of highly oxygenated a-CO$_x$---O:C ratio $\approx 0.6$, according to 
Ref.~\cite{santini_2015}---was
used with $\gamma = N_{\rm C}\gamma_{\rm C}^{\text{high-O}}$ with $ \gamma_{\rm C}^{\text{high-O}} = 19.5$~eV per carbon atom. Unoptimized XPS simulations were performed
on exactly the same structures with $\gamma = 0$~eV. 

\subsubsection{Convergence of GCMC runs}

\begin{figure}[t]
  \centering
  \includegraphics[width=\linewidth]{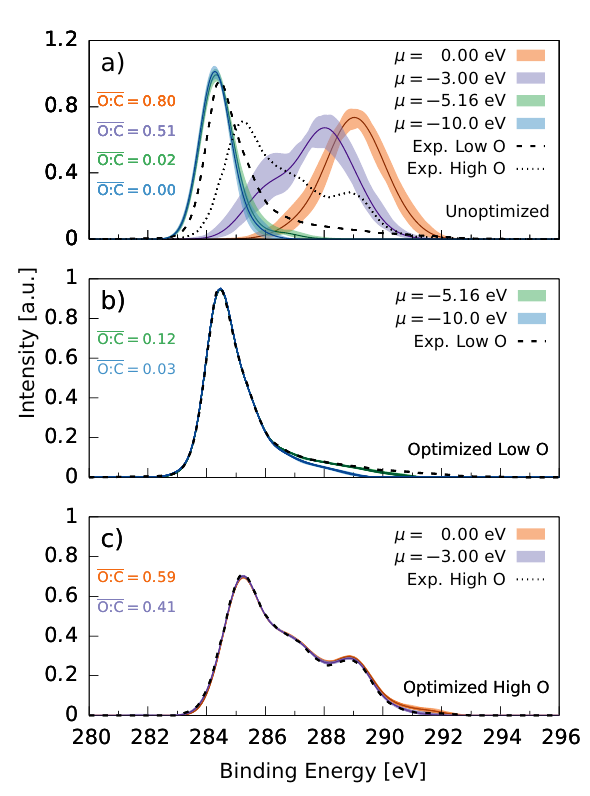}
  \caption{C1s \gls{XPS} spectra of a-CO$_x$ structures (a) without and (b/c) with
    \gls{XPS} optimization. Experimental data taken from
    Ref.~\cite{santini_2015}. b) and c) correspond to optimization of the
    low- and high-oxygen-content spectra, respectively. Filled curves
    denote the maximal extents of the individual runs at the given chemical
    potential; lines give the mean value of the
    spectra.}
  \label{fig:gcmc}
\end{figure}

\begin{figure}[t]
  \centering
  \includegraphics[width=\linewidth]{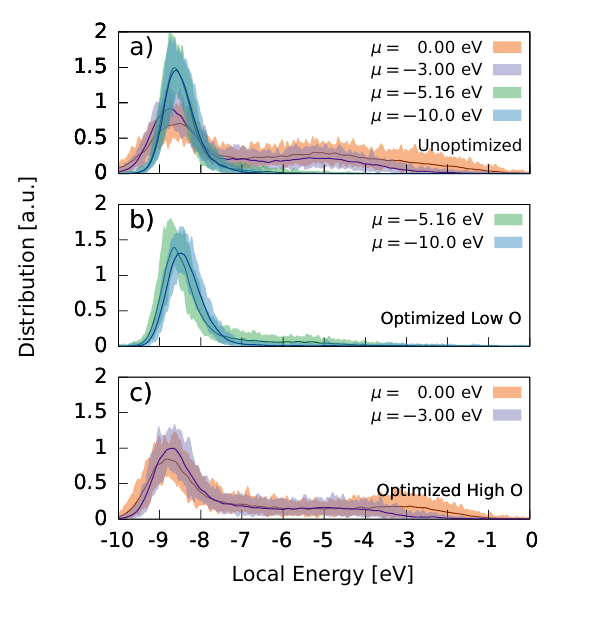}
  \caption{Atomic local-energy distribution of \gls{ACO} structures with and without \gls{XPS} optimization. Filled curves
    denote the maximal extents of the individual runs at the given chemical
    potential; lines give the mean value of the
    distribution. a) Without \gls{XPS} optimization. b) Optimized to low-O
    \gls{ACO} \gls{XPS}. c) Optimized to high-O \gls{ACO} \gls{XPS}. These plots are
    smoothed by Gaussian kernel density estimate.}
  \label{fig:local_energy}
\end{figure}

\fig{fig:gcmc_runs} shows the evolution of
the local energy per atom, the dissimilarity, and the O:C ratio during
the modified \gls{GCMC} runs. The left-hand-side panels show structures
optimized with respect to the low-oxygen-content a-CO$_x$ \gls{XPS}
spectrum, seen in \fig{fig:experimental_xps} a), and the right-hand-side panels
with respect to the high-oxygen-content \gls{XPS} spectrum,
seen in \fig{fig:experimental_xps} b). The color saturation of the lines reflects
the density of the structures, which are in the order of left-to-right, top-to-bottom in Fig.~\ref{fig:structures_bond_angle}: lighter lines correspond to lower density structures, and darker lines to higher densities. Convergence in the energy per atom, dissimilarity and O:C ratio was reached by $2\times 10^5$ steps for all simulations.

Generalized Hamiltonian Monte Carlo initially acted to match the spectra by
increasing oxygen content, after which there was more structural relaxation
and equilibration. The energy per atom increased with oxygen content as the
spectral dissimilarity decreased up to $10^3$ MC steps. After this point,
the deviations of the spectra were small and the local energy decreased. We show in Fig. S6 of the \gls{SI} that the change in the potential dominates that of the change $E_{\rm spectra}$ for configurations with a sufficiently low spectral dissimilarity. The change in potential is generally positive for resultant structures, demonstrating atomic relaxation.

There was a diversity of O:C ratios found at a given chemical potential. The
aptitude for oxygen insertion was highly dependent on the density of the
structure, with those of lower density preferring a greater amount of oxygen.
This is apparent in the bifurcations of the O:C ratio in
\fig{fig:gcmc_runs} c). Structures with greater amounts of oxygen had a higher
energy per atom overall. 

The original analysis of the low-oxygen-content a-C sample found in Santini \emph{et al.}
gave an O:C
ratio of 0.1, which agrees with our low-density \gls{ACO} simulations
of $\mu = -5.16$~eV. The \gls{ACO} results of Santini \emph{et al.},
gave O:C ratios of 0.4--0.8, which are comparable to all our
$\mu = 0$~eV structures and the low-density structures found at $\mu = -3$~eV.
Below we show that, although it is possible to generate
metastable \gls{ACO} structures computationally with similarly high O:C ratios
without \gls{XPS} optimization, the resultant \gls{XPS} spectra are very far from
experiment and the actual O:C
ratio which matches the experimental spectrum is lower.

\begin{figure*}[t]
\includegraphics[width=\linewidth]{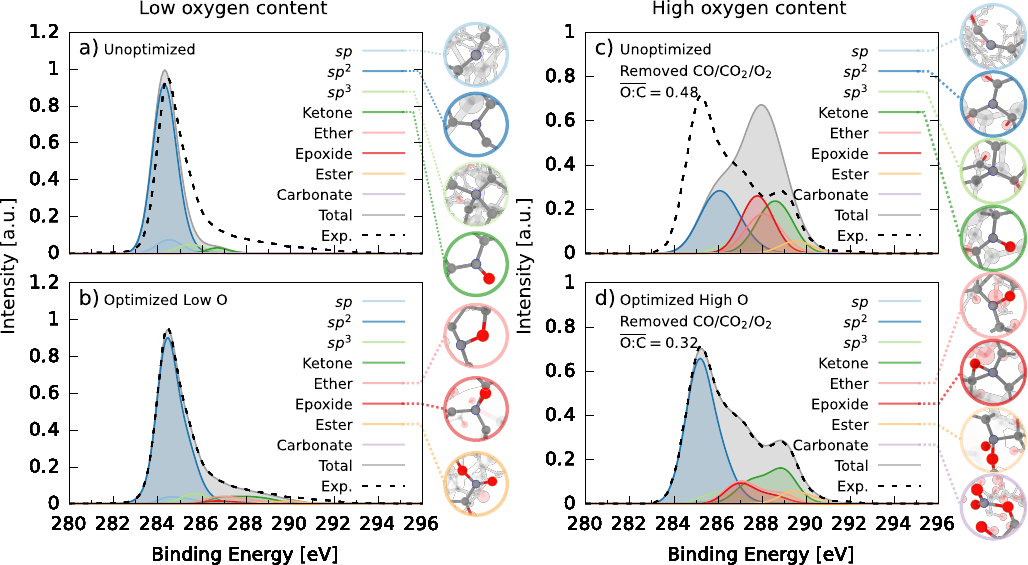}
\caption{Averaged deconvolution of \gls{ACO} structures at low oxygen content with
$\mu = -5.16$~eV (left-hand-side panels) and high oxygen content with $\mu = -3$~eV
and molecular CO$_2$/CO/O$_2$ removed (right-hand-side panels). Images of characteristic
motifs found in the structures are shown on the right of each panel.
a)/c) The unoptimized \gls{XPS} spectra. b)/d) With \gls{XPS} optimization. Contributions
to the high \gls{CEBE} range come from the formation of oxygenated motifs.}
\label{fig:deconvolution_comparison_low}
\label{fig:deconvolution_comparison_high}
\end{figure*}

\subsubsection{Modified vs unmodified GCMC results}

Generalized Hamiltonian Monte Carlo optimization resulted in almost perfect
agreement of the \gls{XPS} spectra for all structures. The \gls{XPS} spectra from the
unoptimized \gls{XPS} and optimized \gls{XPS} runs can be
seen in \fig{fig:gcmc}.
The evolution between a low O-content spectrum and a high O-content spectrum along a single \gls{GCMC} run is shown in Fig.~S1 of the \gls{SI}.
The unoptimized, low-oxygen-content \gls{XPS} spectra had good agreement with experiment.
 The modal peak position of 284.3~eV is at a slightly lower value than the experimental
 284.5~eV, which is expected due to a lack of \gls{CEBE} shifts from high oxygen content
seen in the other samples.
\gls{XPS} optimization resulted in spectral agreement primarily by increasing
oxygen content to reproduce the high \gls{CEBE} tail and causing positive \gls{CEBE} shifts. 
Optimization completely inhibited the large deviations from experiment present in the
unoptimized high-oxygen-content spectra, seen in Fig.~\ref{fig:gcmc} a). The secondary peak
at 289~eV is reproduced well by the optimized spectra, and 
related to the inclusion of oxygenated species~\cite{santini_2015}.
The origin and magnitude of \gls{CEBE} shifts exhibited in the spectra
will be explored in Sec.~\ref{sec:deconvolution_analysis}.

\gls{XPS}-optimized structures had low energies, with local-energy distributions
(\figs{fig:local_energy} b)/c)) comparable to
that of standard \gls{GCMC} (\fig{fig:local_energy} a)). The position of the local-energy
maxima in unoptimized and \gls{XPS}-optimized structures was similar, with only slight
shifts upward in the latter for $\mu = -10$~eV and $\mu = -3$~eV. This suggests that
some carbon environments in these simulations had to be displaced away from lower energy
metastable configurations to fit the experimental \gls{XPS} spectra.
The deviations in peak height can be attributed to oxygen content: greater amounts of
oxygen caused a more pronounced broadening of the local energy distribution.

From these results, we conclude that the inclusion of the spectral penalty term in
\eq{eq:dissimilarity} imposed a sufficient constraint on the Monte Carlo sampling to
promote spectral agreement, while also allowing for the generation of low-energy structures.
The action of this constraint caused the total oxygen content to differ from that of the
unoptimized \gls{XPS} simulations, allowing for an adequate reproduction of the \gls{XPS}
features present in the experimental spectra. The total oxygen content increased for the
low-oxygen \gls{XPS} simulations to obtain the high \gls{CEBE} tails, while oxygen content
decreased for the high chemical potential simulations, to inhibit large sp$^2$ transformations
and sizeable oxygen-induced \gls{CEBE} shifts. For further reference, a comparison between \gls{XPS}-optimized and -unoptimized structures, at
similar oxygen content, is given in the \gls{SI} (Fig.~S2), showing a marked
difference.

The experimental growth process for many a-C materials proceeds through physical vapor deposition. This is a highly energetic, non-equilibrium
process. The products formed will generally be metastable, rather than thermodynamically favorable. Structure generation methods which proceed by sampling traditional partition functions will tend to produce more thermodynamically favorable structures which maximize the entropy---as demonstrated by the melt-quench protocol in Sec.~\ref{sec:melt_quench}. The generalized Hamiltionan approach provides a means to efficiently search the configuration space of the metastable, experimentally viable structures.

\subsection{Motifs and deconvolution}\label{sec:motifs_and_deconvolution}

The deconvolution of the low-oxygen-content spectra,
\fig{fig:deconvolution_comparison_low} a) and b), showed that sp$^2$ carbon motifs dominate
the lower end of the spectra for both unoptimized and optimized simulations.
sp$^2$ composed the majority of the large peak at $\sim$285~eV, similar to the experimental
deconvolution results in Fig.~\ref{fig:experimental_xps} a). The proportions of sp, sp$^2$
and sp$^3$ motifs were similar between optimized and unoptimized simulations.
\gls{XPS} optimization added small amounts of oxygen (in the form of ketone, ether and
epoxide groups) that contributed at higher binding energies than sp$^3$ to account for
the large tail of the experimental \gls{XPS} spectrum.

The effect of \gls{XPS} optimization at high oxygen content was to preserve
a large proportion of sp$^2$ motifs, see \fig{fig:deconvolution_comparison_high} c)/d).
For both the unoptimized \fig{fig:deconvolution_comparison_high} c) and optimized
simulations \fig{fig:deconvolution_comparison_high} d), sp$^2$ motifs gave the largest
contribution to the \gls{XPS} and composed the lower \gls{CEBE} ranges, in addition to sp$^3$.
Many sp$^2$ motifs and all sp motifs present in the initial a-C structures were transformed to
ketone, ether and epoxide groups, which constituted the middle and upper \gls{XPS} range. These
contributions all significantly overlap. Ester and carbonate groups were only found in the high
\gls{CEBE} ranges. \gls{XPS} optimization inhibited the
number of sp$^2$ transformations which could take place, resulting in more
sp$^2$ motifs in comparison to the unoptimized simulations. The oxygenated groups can be
attributed to the secondary peak in the experimental \gls{XPS} spectrum. The \gls{CEBE}
ordering of motif contributions conformed with experimental deconvolution
references~\cite{santini_2015,smith_2016}. 
 
Molecular species CO, CO$_2$ and O$_2$ were formed in these simulations.
The formation of O$_2$ was an artifact of the simulations being performed
at a high chemical potential. The removal of these molecules did not noticeably
impact the resulting XPS spectrum: compare Fig.~\ref{fig:gcmc} c)
and Fig.~\ref{fig:deconvolution_comparison_high} d).
 
\begin{figure}
  \centering
  \includegraphics[width=\linewidth]{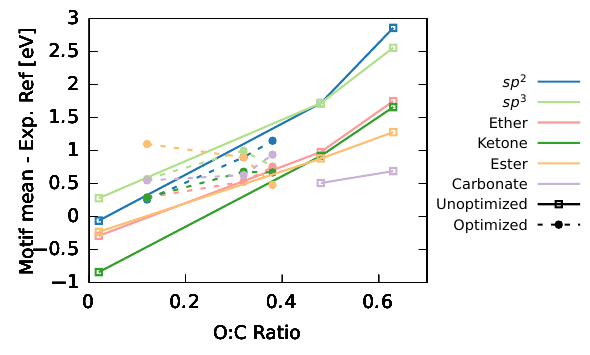}
  \caption{Difference of the mean \gls{CEBE} of a given motif with experimental references subtracted,
  taken from Refs.~\cite{moulder_1992, Haubner2010}, as a function of O:C ratio.
  Experimental references of Moulder \textit{et al.}~\cite{moulder_1992} are taken to be
  the midpoint of the ranges. Unoptimized simulations are the solid lines and optimized
  the dashed. \gls{CEBE} shifts are linear with oxygen content. There were no carbonate
  groups formed at low oxygen content in the unoptimized simulations, hence the lack of
  data in this range. }
  \label{fig:shift_compared_to_exp}
\end{figure}

\glspl{CEBE} increased linearly with bonded oxygen content. The mean \gls{CEBE} motif
contributions at different oxygen contents are compared to experimental references in
Fig.~\ref{fig:shift_compared_to_exp}. The deviations from experimental references come
from at least two factors: 1) The experimental references are molecular: these are bulk simulations;
2) Oxygen is electronegative, which reduces the potential locally, therefore increasing the
\glspl{CEBE} with oxygen content.

\subsection{Discussion}

Our \gls{XPS} optimization method provides a means to efficiently generate experimentally
viable \gls{ACO} structure candidates for different input experimental \gls{XPS} spectra.
Using a dissimilarity measure of the spectra, an efficient search of viable structures
is achieved by a traversal of configuration space that is sensitive to both the total
energy of the structure and spectral agreement. It has distinct advantages over using the
melt-quench \gls{MD} procedure: 1) the \gls{XPS} spectrum of the generated structure
matches that of experiment, even after the removal of extraneous molecules; 2) high-oxygen
content, metastable structures can be generated: the melt-quench procedure does not allow
this due to uncontrolled carbon burning---the generation of CO/CO$_2$.
The inability of unmodified \gls{GCMC} or melt-quench procedures to produce structures
compatible with experimental
spectra is a consequence of the large number of equilibrium atomic configurations which
are possible for amorphous structures. The probability of obtaining experimentally
feasible metastable structures (i.e., those compatible with the experimental synthesis
process) is small without guidance from experimental data.

\gls{XPS} is limited by the fact that structurally dissimilar motifs can sometimes
contribute to the spectral intensity at the same \gls{CEBE}. This is not a computational
artifact, but an intrinsic characteristic of \gls{XPS} as an analytical method, and it
also plagues experimental deconvolution techniques. Thus,
multiple structural candidates can match the chosen \gls{XPS} spectrum, even within the approach
introduced here. This shows that a large configuration space is still available after
restriction by experimental \gls{XPS} spectrum conformation, in the case of amorphous materials.
This has been demonstrated recently by use of a diffusion model to generate a-C structural
candidates which match \gls{XANES} data~\cite{kwon_2023b}. To determine a unique structure,
more experimental data can be included to the multi-objective optimization to restrict the search
in configuration space, as is done in multiple \gls{RMC}
approaches~\cite{opletal_2013,opletal_2017,khadka_2020,farmahini_2013,bacilla_2022}.
The TurboGAP code~\cite{ref_turbogap} already allows for this possibility with regards
to \gls{XRD} and \gls{SAXS} data, and it will be investigated in future work. 

One is not limited to exploring the experimentally constrained configuration space by Monte Carlo techniques. We expect that evolutionary algorithms---such as USPEX \cite{Oganov2006,Oganov2011,Lyakhov2013}---would allow for sufficient sampling of this space to find experimentally viable, low-energy structures.

\section{Conclusion}

Oxygen can induce significant positive \gls{CEBE} shifts in amorphous carbon motifs which
interfere with experimental \gls{XPS} interpretation. These shifts arise from C1s core
electrons experiencing a more negative potential due to negatively charged oxygen in the
local environment, resulting in higher \glspl{CEBE} overall. This effect
renders structurally similar experimental \gls{CEBE} references incomparable,
leading to an erroneous \gls{XPS} interpretation upon deconvolution. 

The effect of these shifts on experimental interpretation is evident in the differences of the high-oxygen content deconvolution from the experimental analysis of Santini \emph{et al.}~\cite{santini_2015},
\fig{fig:experimental_xps} (b), and that of this work, Fig.~\ref{fig:deconvolution_comparison_high}
d). Santini \emph{et al.} suggested that a transformation of sp$^2$ to sp$^3$ carbon takes
place in \gls{ACO} upon a five minute anneal with temperature increasing from 100--500$^\circ$~C.
The sp$^2$ reference energy was not shifted to account for the presence of oxygen, leading the
authors to deduce that all non-oxygenated carbon was of sp$^3$ type from the \gls{XPS} spectrum
deconvolution. Significant numbers of sp$^2$ $\rightarrow$ sp$^3$ transitions did not occur in our
simulations, and we showed that the effect of environmental oxygen is sufficient to explain the presence
of a strongly shifted sp$^2$ peak in \gls{ACO} without the need to invoke the presence of sp$^3$
carbon.

We expect that similar shifts of \glspl{CEBE} occur in materials with other electronegative
species, with positive ions producing the opposite effect: a decrease in \gls{CEBE} with increasing
species content. A thorough investigation of the induced \gls{CEBE} shifts with species content
is necessary to further correct C1s \gls{XPS} interpretation. This can be achieved using the
methodology we have presented in this work.

\section{Outlook}

This work introduces a paradigm by which theory and experiment can be combined to generate
low-energy atomistic structures with \gls{XPS} predictions that conform to experimental
spectra by design. 
This method reconciles experimental and simulated analysis of the structure of materials and,
because of this, it appeals to both the computational scientist---who can have greater confidence
in their results due to experimental agreement---and the experimentalist---who can obtain atomic-scale
information about the structure and potential chemistry of their material.

By imposing agreement of simulated and experimental \gls{XPS} data, new insights into the structure
and composition of real materials can be gained without the inaccuracies apparent in experimental
\gls{XPS} deconvolution techniques, which can give an incorrect picture of a material's constituents.
Furthermore, one is not limited to using one experimental observable: although we have focused here on
\gls{XPS}, the approach can be readily extended to other analytical techniques. Multiple
observables can be used simultaneously to elucidate specific experimental structures. 

Prior approaches used empirical interatomic potentials and were limited to experimental observables
that could be reproduced with simple analytical models. This work leverages the flexibility and
predictive power of atomistic \gls{ML} and paves the way for combining more complex computational
models of experimental observables and \textit{ab-initio}-accurate \glspl{MLP}
to generate structures consistent with experiment.

\section*{Supporting Information}

The Supporting Information accompanying this paper provides miscellaneous additional
information, further tests and benchmarks of our methodology, and technical details
of the calculations: i) evolution of the \gls{XPS} spectrum during a modified
grand-canonical Monte Carlo example run; ii) difference in unoptimized and optimized
grand-canonical Monte Carlo runs for similar oxygen content; iii) variation in modified
grand-canonical Monte Carlo runs with different $\sigma$; iv) technical parameters of
the \gls{DFT} calculations; v) correlation between \gls{CEBE} error and local energy;
vi) demonstration of the magnitude of $E_{\rm pot}$ and $E_{\rm spectra}$ variation.

\begin{acknowledgments}
T.~Z., R.~I. and M.~A.~C. acknowledge financial support from the Research Council of Finland
under grants nos. 321713, 330488, 336304, 355301 and 358050, and computational resources
from CSC (the Finnish IT Center for Science) and Aalto University's Science-IT
Project.
A.~B.~P. acknowledges support from the CASTEP-USER project, funded by the Engineering and Physical Sciences Research Council under the grant agreement EP/W030438/1.
We also thank Prof. Volker~L. Deringer from the University of Oxford
for helpful comments on a draft version of the paper.
\end{acknowledgments}

\def\bibsection{}
\section*{References}

\newpage

\begin{figure}
    \centering
    \includegraphics{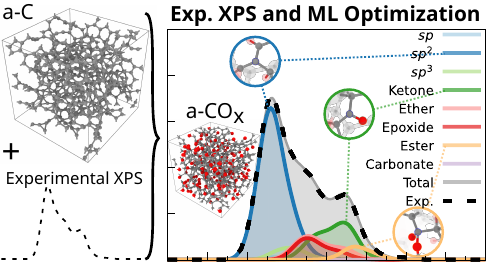}
    \caption{For Table of Contents Only}
    \label{fig:enter-label}
\end{figure}


\begin{thebibliography}{90}%
\makeatletter
\providecommand \@ifxundefined [1]{%
 \@ifx{#1\undefined}
}%
\providecommand \@ifnum [1]{%
 \ifnum #1\expandafter \@firstoftwo
 \else \expandafter \@secondoftwo
 \fi
}%
\providecommand \@ifx [1]{%
 \ifx #1\expandafter \@firstoftwo
 \else \expandafter \@secondoftwo
 \fi
}%
\providecommand \natexlab [1]{#1}%
\providecommand \enquote  [1]{``#1''}%
\providecommand \bibnamefont  [1]{#1}%
\providecommand \bibfnamefont [1]{#1}%
\providecommand \citenamefont [1]{#1}%
\providecommand \href@noop [0]{\@secondoftwo}%
\providecommand \href [0]{\begingroup \@sanitize@url \@href}%
\providecommand \@href[1]{\@@startlink{#1}\@@href}%
\providecommand \@@href[1]{\endgroup#1\@@endlink}%
\providecommand \@sanitize@url [0]{\catcode `\\12\catcode `\$12\catcode `\&12\catcode `\#12\catcode `\^12\catcode `\_12\catcode `\%12\relax}%
\providecommand \@@startlink[1]{}%
\providecommand \@@endlink[0]{}%
\providecommand \url  [0]{\begingroup\@sanitize@url \@url }%
\providecommand \@url [1]{\endgroup\@href {#1}{\urlprefix }}%
\providecommand \urlprefix  [0]{URL }%
\providecommand \Eprint [0]{\href }%
\providecommand \doibase [0]{http://dx.doi.org/}%
\providecommand \selectlanguage [0]{\@gobble}%
\providecommand \bibinfo  [0]{\@secondoftwo}%
\providecommand \bibfield  [0]{\@secondoftwo}%
\providecommand \translation [1]{[#1]}%
\providecommand \BibitemOpen [0]{}%
\providecommand \bibitemStop [0]{}%
\providecommand \bibitemNoStop [0]{.\EOS\space}%
\providecommand \EOS [0]{\spacefactor3000\relax}%
\providecommand \BibitemShut  [1]{\csname bibitem#1\endcsname}%
\let\auto@bib@innerbib\@empty
%</preamble>
\bibitem [{\citenamefont {Behler}\ and\ \citenamefont {Parrinello}(2007)}]{behler_2007}%
  \BibitemOpen
  \bibfield  {author} {\bibinfo {author} {\bibfnamefont {J.}~\bibnamefont {Behler}}\ and\ \bibinfo {author} {\bibfnamefont {M.}~\bibnamefont {Parrinello}},\ }\bibfield  {title} {\enquote {\bibinfo {title} {Generalized neural-network representation of high-dimensional potential-energy surfaces},}\ }\href@noop {} {\bibfield  {journal} {\bibinfo  {journal} {Phys. Rev. Lett.}\ }\textbf {\bibinfo {volume} {98}},\ \bibinfo {pages} {146401} (\bibinfo {year} {2007})}\BibitemShut {NoStop}%
\bibitem [{\citenamefont {Behler}(2017)}]{behler_2017}%
  \BibitemOpen
  \bibfield  {author} {\bibinfo {author} {\bibfnamefont {J.}~\bibnamefont {Behler}},\ }\bibfield  {title} {\enquote {\bibinfo {title} {First principles neural network potentials for reactive simulations of large molecular and condensed systems},}\ }\href@noop {} {\bibfield  {journal} {\bibinfo  {journal} {Angew. Chem. Int. Ed.}\ }\textbf {\bibinfo {volume} {56}},\ \bibinfo {pages} {12828} (\bibinfo {year} {2017})}\BibitemShut {NoStop}%
\bibitem [{\citenamefont {Bart{\'o}k}\ \emph {et~al.}(2010)\citenamefont {Bart{\'o}k}, \citenamefont {Payne}, \citenamefont {Kondor},\ and\ \citenamefont {Cs{\'a}nyi}}]{bartok_2010}%
  \BibitemOpen
  \bibfield  {author} {\bibinfo {author} {\bibfnamefont {A.~P.}\ \bibnamefont {Bart{\'o}k}}, \bibinfo {author} {\bibfnamefont {M.~C.}\ \bibnamefont {Payne}}, \bibinfo {author} {\bibfnamefont {R.}~\bibnamefont {Kondor}}, \ and\ \bibinfo {author} {\bibfnamefont {G.}~\bibnamefont {Cs{\'a}nyi}},\ }\bibfield  {title} {\enquote {\bibinfo {title} {Gaussian approximation potentials: The accuracy of quantum mechanics, without the electrons},}\ }\href@noop {} {\bibfield  {journal} {\bibinfo  {journal} {Phys. Rev. Lett.}\ }\textbf {\bibinfo {volume} {104}},\ \bibinfo {pages} {136403} (\bibinfo {year} {2010})}\BibitemShut {NoStop}%
\bibitem [{\citenamefont {Deringer}\ \emph {et~al.}(2021)\citenamefont {Deringer}, \citenamefont {Bart{\'o}k}, \citenamefont {Bernstein}, \citenamefont {Wilkins}, \citenamefont {Ceriotti},\ and\ \citenamefont {Cs{\'a}nyi}}]{deringer_2021}%
  \BibitemOpen
  \bibfield  {author} {\bibinfo {author} {\bibfnamefont {V.~L.}\ \bibnamefont {Deringer}}, \bibinfo {author} {\bibfnamefont {A.~P.}\ \bibnamefont {Bart{\'o}k}}, \bibinfo {author} {\bibfnamefont {N.}~\bibnamefont {Bernstein}}, \bibinfo {author} {\bibfnamefont {D.~M.}\ \bibnamefont {Wilkins}}, \bibinfo {author} {\bibfnamefont {M.}~\bibnamefont {Ceriotti}}, \ and\ \bibinfo {author} {\bibfnamefont {G.}~\bibnamefont {Cs{\'a}nyi}},\ }\bibfield  {title} {\enquote {\bibinfo {title} {Gaussian process regression for materials and molecules},}\ }\href@noop {} {\bibfield  {journal} {\bibinfo  {journal} {Chem. Rev.}\ }\textbf {\bibinfo {volume} {121}},\ \bibinfo {pages} {10073} (\bibinfo {year} {2021})}\BibitemShut {NoStop}%
\bibitem [{\citenamefont {Butler}\ \emph {et~al.}(2018)\citenamefont {Butler}, \citenamefont {Davies}, \citenamefont {Cartwright}, \citenamefont {Isayev},\ and\ \citenamefont {Walsh}}]{butler_2018}%
  \BibitemOpen
  \bibfield  {author} {\bibinfo {author} {\bibfnamefont {K.~T.}\ \bibnamefont {Butler}}, \bibinfo {author} {\bibfnamefont {D.~W.}\ \bibnamefont {Davies}}, \bibinfo {author} {\bibfnamefont {H.}~\bibnamefont {Cartwright}}, \bibinfo {author} {\bibfnamefont {O.}~\bibnamefont {Isayev}}, \ and\ \bibinfo {author} {\bibfnamefont {A.}~\bibnamefont {Walsh}},\ }\bibfield  {title} {\enquote {\bibinfo {title} {Machine learning for molecular and materials science},}\ }\href@noop {} {\bibfield  {journal} {\bibinfo  {journal} {Nature}\ }\textbf {\bibinfo {volume} {559}},\ \bibinfo {pages} {547} (\bibinfo {year} {2018})}\BibitemShut {NoStop}%
\bibitem [{\citenamefont {Deringer}\ \emph {et~al.}(2019)\citenamefont {Deringer}, \citenamefont {Caro},\ and\ \citenamefont {Cs{\'a}nyi}}]{deringer_2019}%
  \BibitemOpen
  \bibfield  {author} {\bibinfo {author} {\bibfnamefont {V.~L.}\ \bibnamefont {Deringer}}, \bibinfo {author} {\bibfnamefont {M.~A.}\ \bibnamefont {Caro}}, \ and\ \bibinfo {author} {\bibfnamefont {G.}~\bibnamefont {Cs{\'a}nyi}},\ }\bibfield  {title} {\enquote {\bibinfo {title} {Machine learning interatomic potentials as emerging tools for materials science},}\ }\href@noop {} {\bibfield  {journal} {\bibinfo  {journal} {Adv. Mater.}\ }\textbf {\bibinfo {volume} {31}},\ \bibinfo {pages} {1902765} (\bibinfo {year} {2019})}\BibitemShut {NoStop}%
\bibitem [{\citenamefont {Zuo}\ \emph {et~al.}(2020)\citenamefont {Zuo}, \citenamefont {Chen}, \citenamefont {Li}, \citenamefont {Deng}, \citenamefont {Chen}, \citenamefont {Behler}, \citenamefont {Cs{\'a}nyi}, \citenamefont {Shapeev}, \citenamefont {Thompson}, \citenamefont {Wood},\ and\ \citenamefont {Ong}}]{zuo_2020}%
  \BibitemOpen
  \bibfield  {author} {\bibinfo {author} {\bibfnamefont {Y.}~\bibnamefont {Zuo}}, \bibinfo {author} {\bibfnamefont {C.}~\bibnamefont {Chen}}, \bibinfo {author} {\bibfnamefont {X.}~\bibnamefont {Li}}, \bibinfo {author} {\bibfnamefont {Z.}~\bibnamefont {Deng}}, \bibinfo {author} {\bibfnamefont {Y.}~\bibnamefont {Chen}}, \bibinfo {author} {\bibfnamefont {J.}~\bibnamefont {Behler}}, \bibinfo {author} {\bibfnamefont {G.}~\bibnamefont {Cs{\'a}nyi}}, \bibinfo {author} {\bibfnamefont {A.~V.}\ \bibnamefont {Shapeev}}, \bibinfo {author} {\bibfnamefont {A.~P.}\ \bibnamefont {Thompson}}, \bibinfo {author} {\bibfnamefont {M.~A.}\ \bibnamefont {Wood}}, \ and\ \bibinfo {author} {\bibfnamefont {S.~P.}\ \bibnamefont {Ong}},\ }\bibfield  {title} {\enquote {\bibinfo {title} {Performance and cost assessment of machine learning interatomic potentials},}\ }\href@noop {} {\bibfield  {journal} {\bibinfo  {journal} {J. Chem. Phys. A}\ }\textbf {\bibinfo {volume} {124}},\ \bibinfo {pages} {731} (\bibinfo {year} {2020})}\BibitemShut
  {NoStop}%
\bibitem [{\citenamefont {Sch\"utt}\ \emph {et~al.}(2020)\citenamefont {Sch\"utt}, \citenamefont {Chmiela}, \citenamefont {von Lilienfeld}, \citenamefont {Tkatchenko}, \citenamefont {Tsuda},\ and\ \citenamefont {M\"uller}}]{schutt_2020}%
  \BibitemOpen
  \bibfield  {author} {\bibinfo {author} {\bibfnamefont {K.~T.}\ \bibnamefont {Sch\"utt}}, \bibinfo {author} {\bibfnamefont {S.}~\bibnamefont {Chmiela}}, \bibinfo {author} {\bibfnamefont {O.~A.}\ \bibnamefont {von Lilienfeld}}, \bibinfo {author} {\bibfnamefont {A.}~\bibnamefont {Tkatchenko}}, \bibinfo {author} {\bibfnamefont {K.}~\bibnamefont {Tsuda}}, \ and\ \bibinfo {author} {\bibfnamefont {K.-R.}\ \bibnamefont {M\"uller}},\ }\href@noop {} {\emph {\bibinfo {title} {Machine learning meets quantum physics}}},\ \bibinfo {edition} {1st}\ ed.\ (\bibinfo  {publisher} {Springer},\ \bibinfo {address} {Cham, Switzerland},\ \bibinfo {year} {2020})\BibitemShut {NoStop}%
\bibitem [{\citenamefont {Grisafi}\ \emph {et~al.}(2018)\citenamefont {Grisafi}, \citenamefont {Fabrizio}, \citenamefont {Meyer}, \citenamefont {Wilkins}, \citenamefont {Corminboeuf},\ and\ \citenamefont {Ceriotti}}]{grisafi_2018}%
  \BibitemOpen
  \bibfield  {author} {\bibinfo {author} {\bibfnamefont {A.}~\bibnamefont {Grisafi}}, \bibinfo {author} {\bibfnamefont {A.}~\bibnamefont {Fabrizio}}, \bibinfo {author} {\bibfnamefont {B.}~\bibnamefont {Meyer}}, \bibinfo {author} {\bibfnamefont {D.~M.}\ \bibnamefont {Wilkins}}, \bibinfo {author} {\bibfnamefont {C.}~\bibnamefont {Corminboeuf}}, \ and\ \bibinfo {author} {\bibfnamefont {M.}~\bibnamefont {Ceriotti}},\ }\bibfield  {title} {\enquote {\bibinfo {title} {Transferable machine-learning model of the electron density},}\ }\href@noop {} {\bibfield  {journal} {\bibinfo  {journal} {ACS Cent. Sci.}\ }\textbf {\bibinfo {volume} {5}},\ \bibinfo {pages} {57} (\bibinfo {year} {2018})}\BibitemShut {NoStop}%
\bibitem [{\citenamefont {Ben~Mahmoud}\ \emph {et~al.}(2020)\citenamefont {Ben~Mahmoud}, \citenamefont {Anelli}, \citenamefont {Cs{\'a}nyi},\ and\ \citenamefont {Ceriotti}}]{benmahmoud_2020}%
  \BibitemOpen
  \bibfield  {author} {\bibinfo {author} {\bibfnamefont {C.}~\bibnamefont {Ben~Mahmoud}}, \bibinfo {author} {\bibfnamefont {A.}~\bibnamefont {Anelli}}, \bibinfo {author} {\bibfnamefont {G.}~\bibnamefont {Cs{\'a}nyi}}, \ and\ \bibinfo {author} {\bibfnamefont {M.}~\bibnamefont {Ceriotti}},\ }\bibfield  {title} {\enquote {\bibinfo {title} {Learning the electronic density of states in condensed matter},}\ }\href@noop {} {\bibfield  {journal} {\bibinfo  {journal} {Phys. Rev. B}\ }\textbf {\bibinfo {volume} {102}},\ \bibinfo {pages} {235130} (\bibinfo {year} {2020})}\BibitemShut {NoStop}%
\bibitem [{\citenamefont {Gastegger}\ \emph {et~al.}(2020)\citenamefont {Gastegger}, \citenamefont {McSloy}, \citenamefont {Luya}, \citenamefont {Sch{\"u}tt},\ and\ \citenamefont {Maurer}}]{gastegger_2020}%
  \BibitemOpen
  \bibfield  {author} {\bibinfo {author} {\bibfnamefont {M.}~\bibnamefont {Gastegger}}, \bibinfo {author} {\bibfnamefont {A.}~\bibnamefont {McSloy}}, \bibinfo {author} {\bibfnamefont {M.}~\bibnamefont {Luya}}, \bibinfo {author} {\bibfnamefont {K.~T.}\ \bibnamefont {Sch{\"u}tt}}, \ and\ \bibinfo {author} {\bibfnamefont {R.~J.}\ \bibnamefont {Maurer}},\ }\bibfield  {title} {\enquote {\bibinfo {title} {A deep neural network for molecular wave functions in quasi-atomic minimal basis representation},}\ }\href@noop {} {\bibfield  {journal} {\bibinfo  {journal} {J. Chem. Phys.}\ }\textbf {\bibinfo {volume} {153}} (\bibinfo {year} {2020})}\BibitemShut {NoStop}%
\bibitem [{\citenamefont {Bereau}\ \emph {et~al.}(2018)\citenamefont {Bereau}, \citenamefont {DiStasio~Jr}, \citenamefont {Tkatchenko},\ and\ \citenamefont {Von~Lilienfeld}}]{bereau_2018}%
  \BibitemOpen
  \bibfield  {author} {\bibinfo {author} {\bibfnamefont {T.}~\bibnamefont {Bereau}}, \bibinfo {author} {\bibfnamefont {R.~A.}\ \bibnamefont {DiStasio~Jr}}, \bibinfo {author} {\bibfnamefont {A.}~\bibnamefont {Tkatchenko}}, \ and\ \bibinfo {author} {\bibfnamefont {O.~A.}\ \bibnamefont {Von~Lilienfeld}},\ }\bibfield  {title} {\enquote {\bibinfo {title} {Non-covalent interactions across organic and biological subsets of chemical space: Physics-based potentials parametrized from machine learning},}\ }\href@noop {} {\bibfield  {journal} {\bibinfo  {journal} {J. Chem. Phys.}\ }\textbf {\bibinfo {volume} {148}},\ \bibinfo {pages} {241706} (\bibinfo {year} {2018})}\BibitemShut {NoStop}%
\bibitem [{\citenamefont {Muhli}\ \emph {et~al.}(2021)\citenamefont {Muhli}, \citenamefont {Chen}, \citenamefont {Bart\'ok}, \citenamefont {Hern\'andez-Le\'on}, \citenamefont {Cs\'anyi}, \citenamefont {Ala-Nissila},\ and\ \citenamefont {Caro}}]{muhli_2021b}%
  \BibitemOpen
  \bibfield  {author} {\bibinfo {author} {\bibfnamefont {H.}~\bibnamefont {Muhli}}, \bibinfo {author} {\bibfnamefont {X.}~\bibnamefont {Chen}}, \bibinfo {author} {\bibfnamefont {A.~P.}\ \bibnamefont {Bart\'ok}}, \bibinfo {author} {\bibfnamefont {P.}~\bibnamefont {Hern\'andez-Le\'on}}, \bibinfo {author} {\bibfnamefont {G.}~\bibnamefont {Cs\'anyi}}, \bibinfo {author} {\bibfnamefont {T.}~\bibnamefont {Ala-Nissila}}, \ and\ \bibinfo {author} {\bibfnamefont {M.~A.}\ \bibnamefont {Caro}},\ }\bibfield  {title} {\enquote {\bibinfo {title} {Machine learning force fields based on local parametrization of dispersion interactions: Application to the phase diagram of {C$_{60}$}},}\ }\href@noop {} {\bibfield  {journal} {\bibinfo  {journal} {Phys. Rev. B}\ }\textbf {\bibinfo {volume} {104}},\ \bibinfo {pages} {054106} (\bibinfo {year} {2021})}\BibitemShut {NoStop}%
\bibitem [{\citenamefont {Ko}\ \emph {et~al.}(2021)\citenamefont {Ko}, \citenamefont {Finkler}, \citenamefont {Goedecker},\ and\ \citenamefont {Behler}}]{ko_2021}%
  \BibitemOpen
  \bibfield  {author} {\bibinfo {author} {\bibfnamefont {T.~W.}\ \bibnamefont {Ko}}, \bibinfo {author} {\bibfnamefont {J.~A.}\ \bibnamefont {Finkler}}, \bibinfo {author} {\bibfnamefont {S.}~\bibnamefont {Goedecker}}, \ and\ \bibinfo {author} {\bibfnamefont {J.}~\bibnamefont {Behler}},\ }\bibfield  {title} {\enquote {\bibinfo {title} {General-purpose machine learning potentials capturing nonlocal charge transfer},}\ }\href@noop {} {\bibfield  {journal} {\bibinfo  {journal} {Acc. Chem. Res.}\ }\textbf {\bibinfo {volume} {54}},\ \bibinfo {pages} {808} (\bibinfo {year} {2021})}\BibitemShut {NoStop}%
\bibitem [{\citenamefont {Veit}\ \emph {et~al.}(2020)\citenamefont {Veit}, \citenamefont {Wilkins}, \citenamefont {Yang}, \citenamefont {DiStasio~Jr},\ and\ \citenamefont {Ceriotti}}]{veit_2020}%
  \BibitemOpen
  \bibfield  {author} {\bibinfo {author} {\bibfnamefont {M.}~\bibnamefont {Veit}}, \bibinfo {author} {\bibfnamefont {D.~M.}\ \bibnamefont {Wilkins}}, \bibinfo {author} {\bibfnamefont {Y.}~\bibnamefont {Yang}}, \bibinfo {author} {\bibfnamefont {R.~A.}\ \bibnamefont {DiStasio~Jr}}, \ and\ \bibinfo {author} {\bibfnamefont {M.}~\bibnamefont {Ceriotti}},\ }\bibfield  {title} {\enquote {\bibinfo {title} {Predicting molecular dipole moments by combining atomic partial charges and atomic dipoles},}\ }\href@noop {} {\bibfield  {journal} {\bibinfo  {journal} {J. Chem. Phys.}\ }\textbf {\bibinfo {volume} {153}},\ \bibinfo {pages} {024113} (\bibinfo {year} {2020})}\BibitemShut {NoStop}%
\bibitem [{\citenamefont {Staacke}\ \emph {et~al.}(2022)\citenamefont {Staacke}, \citenamefont {Wengert}, \citenamefont {Kunkel}, \citenamefont {Cs\'anyi}, \citenamefont {Reuter},\ and\ \citenamefont {Margraf}}]{staacke_2022}%
  \BibitemOpen
  \bibfield  {author} {\bibinfo {author} {\bibfnamefont {C.~G.}\ \bibnamefont {Staacke}}, \bibinfo {author} {\bibfnamefont {S.}~\bibnamefont {Wengert}}, \bibinfo {author} {\bibfnamefont {C.}~\bibnamefont {Kunkel}}, \bibinfo {author} {\bibfnamefont {G.}~\bibnamefont {Cs\'anyi}}, \bibinfo {author} {\bibfnamefont {K.}~\bibnamefont {Reuter}}, \ and\ \bibinfo {author} {\bibfnamefont {J.~T.}\ \bibnamefont {Margraf}},\ }\bibfield  {title} {\enquote {\bibinfo {title} {Kernel charge equilibration: Efficient and accurate prediction of molecular dipole moments with a machine-learning enhanced electron density model},}\ }\href@noop {} {\bibfield  {journal} {\bibinfo  {journal} {Mach. Learn.: Sci. Technol.}\ }\textbf {\bibinfo {volume} {3}},\ \bibinfo {pages} {015032} (\bibinfo {year} {2022})}\BibitemShut {NoStop}%
\bibitem [{\citenamefont {Caro}\ \emph {et~al.}(2018{\natexlab{a}})\citenamefont {Caro}, \citenamefont {Aarva}, \citenamefont {Deringer}, \citenamefont {Cs\'anyi},\ and\ \citenamefont {Laurila}}]{caro_2018c}%
  \BibitemOpen
  \bibfield  {author} {\bibinfo {author} {\bibfnamefont {M.~A.}\ \bibnamefont {Caro}}, \bibinfo {author} {\bibfnamefont {A.}~\bibnamefont {Aarva}}, \bibinfo {author} {\bibfnamefont {V.~L.}\ \bibnamefont {Deringer}}, \bibinfo {author} {\bibfnamefont {G.}~\bibnamefont {Cs\'anyi}}, \ and\ \bibinfo {author} {\bibfnamefont {T.}~\bibnamefont {Laurila}},\ }\bibfield  {title} {\enquote {\bibinfo {title} {Reactivity of amorphous carbon surfaces: rationalizing the role of structural motifs in functionalization using machine learning},}\ }\href@noop {} {\bibfield  {journal} {\bibinfo  {journal} {Chem. Mater.}\ }\textbf {\bibinfo {volume} {30}},\ \bibinfo {pages} {7446} (\bibinfo {year} {2018}{\natexlab{a}})}\BibitemShut {NoStop}%
\bibitem [{\citenamefont {J{\"a}ger}\ \emph {et~al.}(2018)\citenamefont {J{\"a}ger}, \citenamefont {Morooka}, \citenamefont {Canova}, \citenamefont {Himanen},\ and\ \citenamefont {Foster}}]{jager_2018}%
  \BibitemOpen
  \bibfield  {author} {\bibinfo {author} {\bibfnamefont {M.~O.~J.}\ \bibnamefont {J{\"a}ger}}, \bibinfo {author} {\bibfnamefont {E.~V.}\ \bibnamefont {Morooka}}, \bibinfo {author} {\bibfnamefont {F.~F.}\ \bibnamefont {Canova}}, \bibinfo {author} {\bibfnamefont {L.}~\bibnamefont {Himanen}}, \ and\ \bibinfo {author} {\bibfnamefont {A.~S.}\ \bibnamefont {Foster}},\ }\bibfield  {title} {\enquote {\bibinfo {title} {Machine learning hydrogen adsorption on nanoclusters through structural descriptors},}\ }\href@noop {} {\bibfield  {journal} {\bibinfo  {journal} {npj Comput. Mater.}\ }\textbf {\bibinfo {volume} {4}},\ \bibinfo {pages} {37} (\bibinfo {year} {2018})}\BibitemShut {NoStop}%
\bibitem [{\citenamefont {Ghosh}\ \emph {et~al.}(2019)\citenamefont {Ghosh}, \citenamefont {Stuke}, \citenamefont {Todorovi{\'c}}, \citenamefont {J{\o}rgensen}, \citenamefont {Schmidt}, \citenamefont {Vehtari},\ and\ \citenamefont {Rinke}}]{ghosh_2019}%
  \BibitemOpen
  \bibfield  {author} {\bibinfo {author} {\bibfnamefont {K.}~\bibnamefont {Ghosh}}, \bibinfo {author} {\bibfnamefont {A.}~\bibnamefont {Stuke}}, \bibinfo {author} {\bibfnamefont {M.}~\bibnamefont {Todorovi{\'c}}}, \bibinfo {author} {\bibfnamefont {P.~B.}\ \bibnamefont {J{\o}rgensen}}, \bibinfo {author} {\bibfnamefont {M.~N.}\ \bibnamefont {Schmidt}}, \bibinfo {author} {\bibfnamefont {A.}~\bibnamefont {Vehtari}}, \ and\ \bibinfo {author} {\bibfnamefont {P.}~\bibnamefont {Rinke}},\ }\bibfield  {title} {\enquote {\bibinfo {title} {Deep learning spectroscopy: Neural networks for molecular excitation spectra},}\ }\href@noop {} {\bibfield  {journal} {\bibinfo  {journal} {Adv. Sci.}\ }\textbf {\bibinfo {volume} {6}},\ \bibinfo {pages} {1801367} (\bibinfo {year} {2019})}\BibitemShut {NoStop}%
\bibitem [{\citenamefont {Timoshenko}\ \emph {et~al.}(2017)\citenamefont {Timoshenko}, \citenamefont {Lu}, \citenamefont {Lin},\ and\ \citenamefont {Frenkel}}]{timoshenko_2017}%
  \BibitemOpen
  \bibfield  {author} {\bibinfo {author} {\bibfnamefont {J.}~\bibnamefont {Timoshenko}}, \bibinfo {author} {\bibfnamefont {D.}~\bibnamefont {Lu}}, \bibinfo {author} {\bibfnamefont {Y.}~\bibnamefont {Lin}}, \ and\ \bibinfo {author} {\bibfnamefont {A.~I.}\ \bibnamefont {Frenkel}},\ }\bibfield  {title} {\enquote {\bibinfo {title} {Supervised machine-learning-based determination of three-dimensional structure of metallic nanoparticles},}\ }\href@noop {} {\bibfield  {journal} {\bibinfo  {journal} {J. Phys. Chem. Lett.}\ }\textbf {\bibinfo {volume} {8}},\ \bibinfo {pages} {5091} (\bibinfo {year} {2017})}\BibitemShut {NoStop}%
\bibitem [{\citenamefont {Aarva}\ \emph {et~al.}(2019{\natexlab{a}})\citenamefont {Aarva}, \citenamefont {Deringer}, \citenamefont {Sainio}, \citenamefont {Laurila},\ and\ \citenamefont {Caro}}]{aarva_2019}%
  \BibitemOpen
  \bibfield  {author} {\bibinfo {author} {\bibfnamefont {A.}~\bibnamefont {Aarva}}, \bibinfo {author} {\bibfnamefont {V.~L.}\ \bibnamefont {Deringer}}, \bibinfo {author} {\bibfnamefont {S.}~\bibnamefont {Sainio}}, \bibinfo {author} {\bibfnamefont {T.}~\bibnamefont {Laurila}}, \ and\ \bibinfo {author} {\bibfnamefont {M.~A.}\ \bibnamefont {Caro}},\ }\bibfield  {title} {\enquote {\bibinfo {title} {Understanding {X}-ray spectroscopy of carbonaceous materials by combining experiments, density functional theory and machine learning. part {I}: fingerprint spectra},}\ }\href@noop {} {\bibfield  {journal} {\bibinfo  {journal} {Chem. Mater.}\ }\textbf {\bibinfo {volume} {31}},\ \bibinfo {pages} {9243} (\bibinfo {year} {2019}{\natexlab{a}})}\BibitemShut {NoStop}%
\bibitem [{\citenamefont {Aarva}\ \emph {et~al.}(2019{\natexlab{b}})\citenamefont {Aarva}, \citenamefont {Deringer}, \citenamefont {Sainio}, \citenamefont {Laurila},\ and\ \citenamefont {Caro}}]{aarva_2019b}%
  \BibitemOpen
  \bibfield  {author} {\bibinfo {author} {\bibfnamefont {A.}~\bibnamefont {Aarva}}, \bibinfo {author} {\bibfnamefont {V.~L.}\ \bibnamefont {Deringer}}, \bibinfo {author} {\bibfnamefont {S.}~\bibnamefont {Sainio}}, \bibinfo {author} {\bibfnamefont {T.}~\bibnamefont {Laurila}}, \ and\ \bibinfo {author} {\bibfnamefont {M.~A.}\ \bibnamefont {Caro}},\ }\bibfield  {title} {\enquote {\bibinfo {title} {Understanding {X}-ray spectroscopy of carbonaceous materials by combining experiments, density functional theory and machine learning. part {II}: quantitative fitting of spectra},}\ }\href@noop {} {\bibfield  {journal} {\bibinfo  {journal} {Chem. Mater.}\ }\textbf {\bibinfo {volume} {31}},\ \bibinfo {pages} {9256} (\bibinfo {year} {2019}{\natexlab{b}})}\BibitemShut {NoStop}%
\bibitem [{\citenamefont {Carbone}\ \emph {et~al.}(2020)\citenamefont {Carbone}, \citenamefont {Topsakal}, \citenamefont {Lu},\ and\ \citenamefont {Yoo}}]{carbone_2020}%
  \BibitemOpen
  \bibfield  {author} {\bibinfo {author} {\bibfnamefont {M.~R.}\ \bibnamefont {Carbone}}, \bibinfo {author} {\bibfnamefont {M.}~\bibnamefont {Topsakal}}, \bibinfo {author} {\bibfnamefont {D.}~\bibnamefont {Lu}}, \ and\ \bibinfo {author} {\bibfnamefont {S.}~\bibnamefont {Yoo}},\ }\bibfield  {title} {\enquote {\bibinfo {title} {Machine-learning {X}-ray absorption spectra to quantitative accuracy},}\ }\href@noop {} {\bibfield  {journal} {\bibinfo  {journal} {Phys. Rev. Lett.}\ }\textbf {\bibinfo {volume} {124}},\ \bibinfo {pages} {156401} (\bibinfo {year} {2020})}\BibitemShut {NoStop}%
\bibitem [{\citenamefont {Golze}\ \emph {et~al.}(2022)\citenamefont {Golze}, \citenamefont {Hirvensalo}, \citenamefont {P.}, \citenamefont {Aarva}, \citenamefont {Etula}, \citenamefont {Susi}, \citenamefont {Rinke}, \citenamefont {Laurila},\ and\ \citenamefont {Caro}}]{golze_2022}%
  \BibitemOpen
  \bibfield  {author} {\bibinfo {author} {\bibfnamefont {D.}~\bibnamefont {Golze}}, \bibinfo {author} {\bibfnamefont {M.}~\bibnamefont {Hirvensalo}}, \bibinfo {author} {\bibfnamefont {Hern\'andez-Le\'on}\ \bibnamefont {P.}}, \bibinfo {author} {\bibfnamefont {A.}~\bibnamefont {Aarva}}, \bibinfo {author} {\bibfnamefont {J.}~\bibnamefont {Etula}}, \bibinfo {author} {\bibfnamefont {T.}~\bibnamefont {Susi}}, \bibinfo {author} {\bibfnamefont {P.}~\bibnamefont {Rinke}}, \bibinfo {author} {\bibfnamefont {T.}~\bibnamefont {Laurila}}, \ and\ \bibinfo {author} {\bibfnamefont {M.~A.}\ \bibnamefont {Caro}},\ }\bibfield  {title} {\enquote {\bibinfo {title} {Accurate computational prediction of core-electron binding energies in carbon-based materials: A machine-learning model combining {DFT} and {$GW$}},}\ }\href@noop {} {\bibfield  {journal} {\bibinfo  {journal} {Chem. Mater.}\ }\textbf {\bibinfo {volume} {34}},\ \bibinfo {pages} {6240} (\bibinfo {year} {2022})}\BibitemShut {NoStop}%
\bibitem [{\citenamefont {Kwon}\ \emph {et~al.}(2023{\natexlab{a}})\citenamefont {Kwon}, \citenamefont {Sun}, \citenamefont {Hsu}, \citenamefont {Jeong}, \citenamefont {Aydin}, \citenamefont {Sharma}, \citenamefont {Meng}, \citenamefont {Carbone}, \citenamefont {Chen}, \citenamefont {Lu}, \citenamefont {Wan}, \citenamefont {Nielsen},\ and\ \citenamefont {Pham}}]{kwon_2023}%
  \BibitemOpen
  \bibfield  {author} {\bibinfo {author} {\bibfnamefont {H.}~\bibnamefont {Kwon}}, \bibinfo {author} {\bibfnamefont {W.}~\bibnamefont {Sun}}, \bibinfo {author} {\bibfnamefont {T.}~\bibnamefont {Hsu}}, \bibinfo {author} {\bibfnamefont {W.}~\bibnamefont {Jeong}}, \bibinfo {author} {\bibfnamefont {F.}~\bibnamefont {Aydin}}, \bibinfo {author} {\bibfnamefont {S.}~\bibnamefont {Sharma}}, \bibinfo {author} {\bibfnamefont {F.}~\bibnamefont {Meng}}, \bibinfo {author} {\bibfnamefont {M.~R.}\ \bibnamefont {Carbone}}, \bibinfo {author} {\bibfnamefont {X.}~\bibnamefont {Chen}}, \bibinfo {author} {\bibfnamefont {D.}~\bibnamefont {Lu}}, \bibinfo {author} {\bibfnamefont {L.~F.}\ \bibnamefont {Wan}}, \bibinfo {author} {\bibfnamefont {M.~H.}\ \bibnamefont {Nielsen}}, \ and\ \bibinfo {author} {\bibfnamefont {T.~A.}\ \bibnamefont {Pham}},\ }\bibfield  {title} {\enquote {\bibinfo {title} {Harnessing neural networks for elucidating {X}-ray absorption structure--spectrum relationships in amorphous carbon},}\ }\href@noop {}
  {\bibfield  {journal} {\bibinfo  {journal} {J. Phys. Chem. C}\ }\textbf {\bibinfo {volume} {127}},\ \bibinfo {pages} {16473} (\bibinfo {year} {2023}{\natexlab{a}})}\BibitemShut {NoStop}%
\bibitem [{\citenamefont {Kotobi}\ \emph {et~al.}(2023)\citenamefont {Kotobi}, \citenamefont {Singh}, \citenamefont {H\"oche}, \citenamefont {Bari}, \citenamefont {Mei{\ss}ner},\ and\ \citenamefont {Bande}}]{kotobi_2023}%
  \BibitemOpen
  \bibfield  {author} {\bibinfo {author} {\bibfnamefont {A.}~\bibnamefont {Kotobi}}, \bibinfo {author} {\bibfnamefont {K.}~\bibnamefont {Singh}}, \bibinfo {author} {\bibfnamefont {D.}~\bibnamefont {H\"oche}}, \bibinfo {author} {\bibfnamefont {S.}~\bibnamefont {Bari}}, \bibinfo {author} {\bibfnamefont {R.~H.}\ \bibnamefont {Mei{\ss}ner}}, \ and\ \bibinfo {author} {\bibfnamefont {A.}~\bibnamefont {Bande}},\ }\bibfield  {title} {\enquote {\bibinfo {title} {Integrating explainability into graph neural network models for the prediction of {X}-ray absorption spectra},}\ }\href@noop {} {\bibfield  {journal} {\bibinfo  {journal} {J. Am. Chem. Soc.}\ }\textbf {\bibinfo {volume} {145}},\ \bibinfo {pages} {22584} (\bibinfo {year} {2023})}\BibitemShut {NoStop}%
\bibitem [{\citenamefont {Chen}\ \emph {et~al.}(2023)\citenamefont {Chen}, \citenamefont {Chen}, \citenamefont {Hwang}, \citenamefont {Davis}, \citenamefont {Yang}, \citenamefont {Sun}, \citenamefont {Ong},\ and\ \citenamefont {Chan}}]{chen_2023}%
  \BibitemOpen
  \bibfield  {author} {\bibinfo {author} {\bibfnamefont {Y.}~\bibnamefont {Chen}}, \bibinfo {author} {\bibfnamefont {C.}~\bibnamefont {Chen}}, \bibinfo {author} {\bibfnamefont {I.}~\bibnamefont {Hwang}}, \bibinfo {author} {\bibfnamefont {M.~J.}\ \bibnamefont {Davis}}, \bibinfo {author} {\bibfnamefont {W.}~\bibnamefont {Yang}}, \bibinfo {author} {\bibfnamefont {C.}~\bibnamefont {Sun}}, \bibinfo {author} {\bibfnamefont {S.~P.}\ \bibnamefont {Ong}}, \ and\ \bibinfo {author} {\bibfnamefont {M.~K.~Y.}\ \bibnamefont {Chan}},\ }\bibfield  {title} {\enquote {\bibinfo {title} {Robust machine learning inference from {X}-ray absorption near edge spectra through featurization},}\ }\href@noop {} {\bibfield  {journal} {\bibinfo  {journal} {arXiv:2310.07049}\ } (\bibinfo {year} {2023})}\BibitemShut {NoStop}%
\bibitem [{\citenamefont {Schwenker}\ \emph {et~al.}(2022)\citenamefont {Schwenker}, \citenamefont {Kolluru}, \citenamefont {Guo}, \citenamefont {Zhang}, \citenamefont {Hu}, \citenamefont {Li}, \citenamefont {Paul}, \citenamefont {Hersam}, \citenamefont {Dravid}, \citenamefont {Klie}, \citenamefont {Guest},\ and\ \citenamefont {Chan}}]{schwenker_2022}%
  \BibitemOpen
  \bibfield  {author} {\bibinfo {author} {\bibfnamefont {E.}~\bibnamefont {Schwenker}}, \bibinfo {author} {\bibfnamefont {V.~S.~C.}\ \bibnamefont {Kolluru}}, \bibinfo {author} {\bibfnamefont {J.}~\bibnamefont {Guo}}, \bibinfo {author} {\bibfnamefont {R.}~\bibnamefont {Zhang}}, \bibinfo {author} {\bibfnamefont {X.}~\bibnamefont {Hu}}, \bibinfo {author} {\bibfnamefont {Q.}~\bibnamefont {Li}}, \bibinfo {author} {\bibfnamefont {J.~T.}\ \bibnamefont {Paul}}, \bibinfo {author} {\bibfnamefont {M.~C.}\ \bibnamefont {Hersam}}, \bibinfo {author} {\bibfnamefont {V.~P.}\ \bibnamefont {Dravid}}, \bibinfo {author} {\bibfnamefont {R.}~\bibnamefont {Klie}}, \bibinfo {author} {\bibfnamefont {J.~R.}\ \bibnamefont {Guest}}, \ and\ \bibinfo {author} {\bibfnamefont {M.~K.~Y.}\ \bibnamefont {Chan}},\ }\bibfield  {title} {\enquote {\bibinfo {title} {Ingrained: An automated framework for fusing atomic-scale image simulations into experiments},}\ }\href@noop {} {\bibfield  {journal} {\bibinfo  {journal} {Small}\ }\textbf {\bibinfo
  {volume} {18}},\ \bibinfo {pages} {2102960} (\bibinfo {year} {2022})}\BibitemShut {NoStop}%
\bibitem [{\citenamefont {Unruh}\ \emph {et~al.}(2022)\citenamefont {Unruh}, \citenamefont {Kolluru}, \citenamefont {Baskaran}, \citenamefont {Chen},\ and\ \citenamefont {Chan}}]{unruh_2022}%
  \BibitemOpen
  \bibfield  {author} {\bibinfo {author} {\bibfnamefont {D.}~\bibnamefont {Unruh}}, \bibinfo {author} {\bibfnamefont {V.~S.~C.}\ \bibnamefont {Kolluru}}, \bibinfo {author} {\bibfnamefont {A.}~\bibnamefont {Baskaran}}, \bibinfo {author} {\bibfnamefont {Y.}~\bibnamefont {Chen}}, \ and\ \bibinfo {author} {\bibfnamefont {M.~K.~Y.}\ \bibnamefont {Chan}},\ }\bibfield  {title} {\enquote {\bibinfo {title} {Theory+ {AI/ML} for microscopy and spectroscopy: Challenges and opportunities},}\ }\href@noop {} {\bibfield  {journal} {\bibinfo  {journal} {MRS Bull.}\ }\textbf {\bibinfo {volume} {47}},\ \bibinfo {pages} {1024} (\bibinfo {year} {2022})}\BibitemShut {NoStop}%
\bibitem [{\citenamefont {Kwon}\ \emph {et~al.}(2023{\natexlab{b}})\citenamefont {Kwon}, \citenamefont {Hsu}, \citenamefont {Sun}, \citenamefont {Jeong}, \citenamefont {Aydin}, \citenamefont {Chapman}, \citenamefont {Chen}, \citenamefont {Carbone}, \citenamefont {Lu}, \citenamefont {Zhou},\ and\ \citenamefont {Pham}}]{kwon_2023b}%
  \BibitemOpen
  \bibfield  {author} {\bibinfo {author} {\bibfnamefont {H.}~\bibnamefont {Kwon}}, \bibinfo {author} {\bibfnamefont {T.}~\bibnamefont {Hsu}}, \bibinfo {author} {\bibfnamefont {W.}~\bibnamefont {Sun}}, \bibinfo {author} {\bibfnamefont {W.}~\bibnamefont {Jeong}}, \bibinfo {author} {\bibfnamefont {F.}~\bibnamefont {Aydin}}, \bibinfo {author} {\bibfnamefont {J.}~\bibnamefont {Chapman}}, \bibinfo {author} {\bibfnamefont {X.}~\bibnamefont {Chen}}, \bibinfo {author} {\bibfnamefont {M.~R.}\ \bibnamefont {Carbone}}, \bibinfo {author} {\bibfnamefont {D.}~\bibnamefont {Lu}}, \bibinfo {author} {\bibfnamefont {F.}~\bibnamefont {Zhou}}, \ and\ \bibinfo {author} {\bibfnamefont {T.~A.}\ \bibnamefont {Pham}},\ }\bibfield  {title} {\enquote {\bibinfo {title} {Spectroscopy-guided discovery of three-dimensional structures of disordered materials with diffusion models},}\ }\href@noop {} {\bibfield  {journal} {\bibinfo  {journal} {arXiv:2312.05472}\ } (\bibinfo {year} {2023}{\natexlab{b}})}\BibitemShut {NoStop}%
\bibitem [{\citenamefont {Santini}\ \emph {et~al.}(2015)\citenamefont {Santini}, \citenamefont {Sebastian}, \citenamefont {Marchiori}, \citenamefont {Jonnalagadda}, \citenamefont {Dellmann}, \citenamefont {Koelmans}, \citenamefont {Rossell}, \citenamefont {Rossel},\ and\ \citenamefont {Eleftheriou}}]{santini_2015}%
  \BibitemOpen
  \bibfield  {author} {\bibinfo {author} {\bibfnamefont {C.~A.}\ \bibnamefont {Santini}}, \bibinfo {author} {\bibfnamefont {A.}~\bibnamefont {Sebastian}}, \bibinfo {author} {\bibfnamefont {C.}~\bibnamefont {Marchiori}}, \bibinfo {author} {\bibfnamefont {V.~P.}\ \bibnamefont {Jonnalagadda}}, \bibinfo {author} {\bibfnamefont {L.}~\bibnamefont {Dellmann}}, \bibinfo {author} {\bibfnamefont {W.~W.}\ \bibnamefont {Koelmans}}, \bibinfo {author} {\bibfnamefont {M.~D.}\ \bibnamefont {Rossell}}, \bibinfo {author} {\bibfnamefont {C.~P.}\ \bibnamefont {Rossel}}, \ and\ \bibinfo {author} {\bibfnamefont {E.}~\bibnamefont {Eleftheriou}},\ }\bibfield  {title} {\enquote {\bibinfo {title} {Oxygenated amorphous carbon for resistive memory applications},}\ }\href@noop {} {\bibfield  {journal} {\bibinfo  {journal} {Nat. Commun.}\ }\textbf {\bibinfo {volume} {6}},\ \bibinfo {pages} {1} (\bibinfo {year} {2015})}\BibitemShut {NoStop}%
\bibitem [{\citenamefont {Bachmann}\ \emph {et~al.}(2017)\citenamefont {Bachmann}, \citenamefont {Koelmans}, \citenamefont {Jonnalagadda}, \citenamefont {Le~Gallo}, \citenamefont {Santini}, \citenamefont {Sebastian}, \citenamefont {Eleftheriou}, \citenamefont {Craciun},\ and\ \citenamefont {Wright}}]{bachmann_2017}%
  \BibitemOpen
  \bibfield  {author} {\bibinfo {author} {\bibfnamefont {T.~A.}\ \bibnamefont {Bachmann}}, \bibinfo {author} {\bibfnamefont {W.~W.}\ \bibnamefont {Koelmans}}, \bibinfo {author} {\bibfnamefont {V.~P.}\ \bibnamefont {Jonnalagadda}}, \bibinfo {author} {\bibfnamefont {M.}~\bibnamefont {Le~Gallo}}, \bibinfo {author} {\bibfnamefont {C.~A.}\ \bibnamefont {Santini}}, \bibinfo {author} {\bibfnamefont {A.}~\bibnamefont {Sebastian}}, \bibinfo {author} {\bibfnamefont {E.}~\bibnamefont {Eleftheriou}}, \bibinfo {author} {\bibfnamefont {M.~F.}\ \bibnamefont {Craciun}}, \ and\ \bibinfo {author} {\bibfnamefont {C.~D.}\ \bibnamefont {Wright}},\ }\bibfield  {title} {\enquote {\bibinfo {title} {Memristive effects in oxygenated amorphous carbon nanodevices},}\ }\href@noop {} {\bibfield  {journal} {\bibinfo  {journal} {Nanotechnology}\ }\textbf {\bibinfo {volume} {29}},\ \bibinfo {pages} {035201} (\bibinfo {year} {2017})}\BibitemShut {NoStop}%
\bibitem [{\citenamefont {Palom{\"a}ki}\ \emph {et~al.}(2016)\citenamefont {Palom{\"a}ki}, \citenamefont {Wester}, \citenamefont {Johansson}, \citenamefont {Laitinen}, \citenamefont {Jiang}, \citenamefont {Arstila}, \citenamefont {Sajavaara}, \citenamefont {Han}, \citenamefont {Koskinen},\ and\ \citenamefont {Laurila}}]{palomaki_2016}%
  \BibitemOpen
  \bibfield  {author} {\bibinfo {author} {\bibfnamefont {T.}~\bibnamefont {Palom{\"a}ki}}, \bibinfo {author} {\bibfnamefont {N.}~\bibnamefont {Wester}}, \bibinfo {author} {\bibfnamefont {L.-S.}\ \bibnamefont {Johansson}}, \bibinfo {author} {\bibfnamefont {M.}~\bibnamefont {Laitinen}}, \bibinfo {author} {\bibfnamefont {H.}~\bibnamefont {Jiang}}, \bibinfo {author} {\bibfnamefont {K.}~\bibnamefont {Arstila}}, \bibinfo {author} {\bibfnamefont {T.}~\bibnamefont {Sajavaara}}, \bibinfo {author} {\bibfnamefont {J.~G.}\ \bibnamefont {Han}}, \bibinfo {author} {\bibfnamefont {J.}~\bibnamefont {Koskinen}}, \ and\ \bibinfo {author} {\bibfnamefont {T.}~\bibnamefont {Laurila}},\ }\bibfield  {title} {\enquote {\bibinfo {title} {Characterization and electrochemical properties of oxygenated amorphous carbon ({a-C}) films},}\ }\href@noop {} {\bibfield  {journal} {\bibinfo  {journal} {Electrochim. Acta}\ }\textbf {\bibinfo {volume} {220}},\ \bibinfo {pages} {137} (\bibinfo {year} {2016})}\BibitemShut {NoStop}%
\bibitem [{\citenamefont {Qasim}\ \emph {et~al.}(2019)\citenamefont {Qasim}, \citenamefont {Ruan}, \citenamefont {Fu}, \citenamefont {Ali}, \citenamefont {Mehrjou}, \citenamefont {Wu}, \citenamefont {Liu}, \citenamefont {Wu},\ and\ \citenamefont {Chu}}]{qasim_2019}%
  \BibitemOpen
  \bibfield  {author} {\bibinfo {author} {\bibfnamefont {A.~M.}\ \bibnamefont {Qasim}}, \bibinfo {author} {\bibfnamefont {Q.}~\bibnamefont {Ruan}}, \bibinfo {author} {\bibfnamefont {R.~K.~Y.}\ \bibnamefont {Fu}}, \bibinfo {author} {\bibfnamefont {F.}~\bibnamefont {Ali}}, \bibinfo {author} {\bibfnamefont {B.}~\bibnamefont {Mehrjou}}, \bibinfo {author} {\bibfnamefont {H.}~\bibnamefont {Wu}}, \bibinfo {author} {\bibfnamefont {L.}~\bibnamefont {Liu}}, \bibinfo {author} {\bibfnamefont {Z.}~\bibnamefont {Wu}}, \ and\ \bibinfo {author} {\bibfnamefont {P.~K.}\ \bibnamefont {Chu}},\ }\bibfield  {title} {\enquote {\bibinfo {title} {Enhanced oxygen-induced properties of bulk oxygenated amorphous carbon films deposited with an anode layer ion source},}\ }\href@noop {} {\bibfield  {journal} {\bibinfo  {journal} {Vacuum}\ }\textbf {\bibinfo {volume} {169}},\ \bibinfo {pages} {108915} (\bibinfo {year} {2019})}\BibitemShut {NoStop}%
\bibitem [{\citenamefont {Marchon}\ \emph {et~al.}(1990)\citenamefont {Marchon}, \citenamefont {Heiman},\ and\ \citenamefont {Khan}}]{marchon_1990}%
  \BibitemOpen
  \bibfield  {author} {\bibinfo {author} {\bibfnamefont {B.}~\bibnamefont {Marchon}}, \bibinfo {author} {\bibfnamefont {N.}~\bibnamefont {Heiman}}, \ and\ \bibinfo {author} {\bibfnamefont {M.~R.}\ \bibnamefont {Khan}},\ }\bibfield  {title} {\enquote {\bibinfo {title} {Evidence for tribochemical wear on amorphous carbon thin films},}\ }\href@noop {} {\bibfield  {journal} {\bibinfo  {journal} {IEEE Trans. Mag.}\ }\textbf {\bibinfo {volume} {26}},\ \bibinfo {pages} {168} (\bibinfo {year} {1990})}\BibitemShut {NoStop}%
\bibitem [{\citenamefont {Moulder}\ \emph {et~al.}(1992)\citenamefont {Moulder}, \citenamefont {Stickle}, \citenamefont {Sobol},\ and\ \citenamefont {Bomben}}]{moulder_1992}%
  \BibitemOpen
  \bibfield  {author} {\bibinfo {author} {\bibfnamefont {J.~F.}\ \bibnamefont {Moulder}}, \bibinfo {author} {\bibfnamefont {W.~F.}\ \bibnamefont {Stickle}}, \bibinfo {author} {\bibfnamefont {P.~E.}\ \bibnamefont {Sobol}}, \ and\ \bibinfo {author} {\bibfnamefont {K.~D.}\ \bibnamefont {Bomben}},\ }\href@noop {} {\emph {\bibinfo {title} {Handbook of {X}-ray photoelectron spectroscopy}}}\ (\bibinfo  {publisher} {Perkin-Elmer Corporation},\ \bibinfo {address} {Eden Prairie},\ \bibinfo {year} {1992})\BibitemShut {NoStop}%
\bibitem [{\citenamefont {Major}\ \emph {et~al.}(2020)\citenamefont {Major}, \citenamefont {Fairley}, \citenamefont {Sherwood}, \citenamefont {Linford}, \citenamefont {Terry}, \citenamefont {Fernandez},\ and\ \citenamefont {Artyushkova}}]{major_2020}%
  \BibitemOpen
  \bibfield  {author} {\bibinfo {author} {\bibfnamefont {G.~H.}\ \bibnamefont {Major}}, \bibinfo {author} {\bibfnamefont {N.}~\bibnamefont {Fairley}}, \bibinfo {author} {\bibfnamefont {P.}~\bibnamefont {Sherwood}}, \bibinfo {author} {\bibfnamefont {M.~R.}\ \bibnamefont {Linford}}, \bibinfo {author} {\bibfnamefont {J.}~\bibnamefont {Terry}}, \bibinfo {author} {\bibfnamefont {V.}~\bibnamefont {Fernandez}}, \ and\ \bibinfo {author} {\bibfnamefont {K.}~\bibnamefont {Artyushkova}},\ }\bibfield  {title} {\enquote {\bibinfo {title} {Practical guide for curve fitting in x-ray photoelectron spectroscopy},}\ }\href@noop {} {\bibfield  {journal} {\bibinfo  {journal} {J. Vac. Sci. Technol. A}\ }\textbf {\bibinfo {volume} {38}},\ \bibinfo {pages} {061203} (\bibinfo {year} {2020})}\BibitemShut {NoStop}%
\bibitem [{\citenamefont {Gengenbach}\ \emph {et~al.}(2021)\citenamefont {Gengenbach}, \citenamefont {Major}, \citenamefont {Linford},\ and\ \citenamefont {Easton}}]{gengenbach_2021}%
  \BibitemOpen
  \bibfield  {author} {\bibinfo {author} {\bibfnamefont {T.~R.}\ \bibnamefont {Gengenbach}}, \bibinfo {author} {\bibfnamefont {G.~H.}\ \bibnamefont {Major}}, \bibinfo {author} {\bibfnamefont {M.~R.}\ \bibnamefont {Linford}}, \ and\ \bibinfo {author} {\bibfnamefont {C.~D.}\ \bibnamefont {Easton}},\ }\bibfield  {title} {\enquote {\bibinfo {title} {Practical guides for x-ray photoelectron spectroscopy ({XPS}): Interpreting the carbon 1s spectrum},}\ }\href@noop {} {\bibfield  {journal} {\bibinfo  {journal} {J. Vac. Sci. Technol. A}\ }\textbf {\bibinfo {volume} {39}},\ \bibinfo {pages} {013204} (\bibinfo {year} {2021})}\BibitemShut {NoStop}%
\bibitem [{\citenamefont {Bagus}\ \emph {et~al.}(2023)\citenamefont {Bagus}, \citenamefont {Nelin},\ and\ \citenamefont {Brundle}}]{bagus_2023}%
  \BibitemOpen
  \bibfield  {author} {\bibinfo {author} {\bibfnamefont {P.~S.}\ \bibnamefont {Bagus}}, \bibinfo {author} {\bibfnamefont {C.~J.}\ \bibnamefont {Nelin}}, \ and\ \bibinfo {author} {\bibfnamefont {C.~R.}\ \bibnamefont {Brundle}},\ }\bibfield  {title} {\enquote {\bibinfo {title} {Chemical significance of x-ray photoelectron spectroscopy binding energy shifts: A perspective},}\ }\href@noop {} {\bibfield  {journal} {\bibinfo  {journal} {J. Vac. Sci. Technol. A}\ }\textbf {\bibinfo {volume} {41}},\ \bibinfo {pages} {068501} (\bibinfo {year} {2023})}\BibitemShut {NoStop}%
\bibitem [{\citenamefont {Shirley}(1972)}]{Shirley1972}%
  \BibitemOpen
  \bibfield  {author} {\bibinfo {author} {\bibfnamefont {D.~A.}\ \bibnamefont {Shirley}},\ }\bibfield  {title} {\enquote {\bibinfo {title} {High-{{Resolution X-Ray Photoemission Spectrum}} of the {{Valence Bands}} of {{Gold}}},}\ }\href {\doibase 10.1103/PhysRevB.5.4709} {\bibfield  {journal} {\bibinfo  {journal} {Physical Review B}\ }\textbf {\bibinfo {volume} {5}},\ \bibinfo {pages} {4709--4714} (\bibinfo {year} {1972})}\BibitemShut {NoStop}%
\bibitem [{\citenamefont {Nagareddy}\ \emph {et~al.}(2018)\citenamefont {Nagareddy}, \citenamefont {Ott}, \citenamefont {Dou}, \citenamefont {Tsvetkova}, \citenamefont {Sandulov}, \citenamefont {Craciun}, \citenamefont {Ferrari},\ and\ \citenamefont {Wright}}]{nagareddy_2018}%
  \BibitemOpen
  \bibfield  {author} {\bibinfo {author} {\bibfnamefont {V.~K.}\ \bibnamefont {Nagareddy}}, \bibinfo {author} {\bibfnamefont {A.~K.}\ \bibnamefont {Ott}}, \bibinfo {author} {\bibfnamefont {C.}~\bibnamefont {Dou}}, \bibinfo {author} {\bibfnamefont {T.}~\bibnamefont {Tsvetkova}}, \bibinfo {author} {\bibfnamefont {M.}~\bibnamefont {Sandulov}}, \bibinfo {author} {\bibfnamefont {M.~F.}\ \bibnamefont {Craciun}}, \bibinfo {author} {\bibfnamefont {A.~C.}\ \bibnamefont {Ferrari}}, \ and\ \bibinfo {author} {\bibfnamefont {C.~D.}\ \bibnamefont {Wright}},\ }\bibfield  {title} {\enquote {\bibinfo {title} {The effect of nitrogen implantation on resistive switching of tetrahedral amorphous carbon films},}\ }\href@noop {} {\bibfield  {journal} {\bibinfo  {journal} {Diam. Relat. Mater.}\ }\textbf {\bibinfo {volume} {87}},\ \bibinfo {pages} {90} (\bibinfo {year} {2018})}\BibitemShut {NoStop}%
\bibitem [{\citenamefont {Smith}\ \emph {et~al.}(2016)\citenamefont {Smith}, \citenamefont {Scudiero}, \citenamefont {Espinal}, \citenamefont {McEwen},\ and\ \citenamefont {Garcia-Perez}}]{smith_2016}%
  \BibitemOpen
  \bibfield  {author} {\bibinfo {author} {\bibfnamefont {M.}~\bibnamefont {Smith}}, \bibinfo {author} {\bibfnamefont {L.}~\bibnamefont {Scudiero}}, \bibinfo {author} {\bibfnamefont {J.}~\bibnamefont {Espinal}}, \bibinfo {author} {\bibfnamefont {J.-S.}\ \bibnamefont {McEwen}}, \ and\ \bibinfo {author} {\bibfnamefont {M.}~\bibnamefont {Garcia-Perez}},\ }\bibfield  {title} {\enquote {\bibinfo {title} {Improving the deconvolution and interpretation of {XPS} spectra from chars by ab initio calculations},}\ }\href@noop {} {\bibfield  {journal} {\bibinfo  {journal} {Carbon}\ }\textbf {\bibinfo {volume} {110}},\ \bibinfo {pages} {155} (\bibinfo {year} {2016})}\BibitemShut {NoStop}%
\bibitem [{\citenamefont {Dyson}(2004)}]{dyson_2004}%
  \BibitemOpen
  \bibfield  {author} {\bibinfo {author} {\bibfnamefont {F.}~\bibnamefont {Dyson}},\ }\bibfield  {title} {\enquote {\bibinfo {title} {A meeting with {Enrico Fermi}},}\ }\href@noop {} {\bibfield  {journal} {\bibinfo  {journal} {Nature}\ }\textbf {\bibinfo {volume} {427}},\ \bibinfo {pages} {297} (\bibinfo {year} {2004})}\BibitemShut {NoStop}%
\bibitem [{\citenamefont {McGreevy}\ and\ \citenamefont {Pusztai}(1988)}]{McGreevy1988}%
  \BibitemOpen
  \bibfield  {author} {\bibinfo {author} {\bibfnamefont {R.~L.}\ \bibnamefont {McGreevy}}\ and\ \bibinfo {author} {\bibfnamefont {L.}~\bibnamefont {Pusztai}},\ }\bibfield  {title} {\enquote {\bibinfo {title} {Reverse {{Monte Carlo Simulation}}: {{A New Technique}} for the {{Determination}} of {{Disordered Structures}}},}\ }\href {\doibase 10.1080/08927028808080958} {\bibfield  {journal} {\bibinfo  {journal} {Molecular Simulation}\ }\textbf {\bibinfo {volume} {1}},\ \bibinfo {pages} {359--367} (\bibinfo {year} {1988})}\BibitemShut {NoStop}%
\bibitem [{\citenamefont {Biswas}\ \emph {et~al.}(2004)\citenamefont {Biswas}, \citenamefont {Atta-Fynn},\ and\ \citenamefont {Drabold}}]{biswas_2004}%
  \BibitemOpen
  \bibfield  {author} {\bibinfo {author} {\bibfnamefont {P.}~\bibnamefont {Biswas}}, \bibinfo {author} {\bibfnamefont {R.}~\bibnamefont {Atta-Fynn}}, \ and\ \bibinfo {author} {\bibfnamefont {D.~A.}\ \bibnamefont {Drabold}},\ }\bibfield  {title} {\enquote {\bibinfo {title} {Reverse {Monte Carlo} modeling of amorphous silicon},}\ }\href@noop {} {\bibfield  {journal} {\bibinfo  {journal} {Phys. Rev. B}\ }\textbf {\bibinfo {volume} {69}},\ \bibinfo {pages} {195207} (\bibinfo {year} {2004})}\BibitemShut {NoStop}%
\bibitem [{\citenamefont {Limbu}\ \emph {et~al.}(2020)\citenamefont {Limbu}, \citenamefont {Elliott}, \citenamefont {Atta-Fynn},\ and\ \citenamefont {Biswas}}]{limbu_2020}%
  \BibitemOpen
  \bibfield  {author} {\bibinfo {author} {\bibfnamefont {K.}~\bibnamefont {Limbu}, \bibfnamefont {D}}, \bibinfo {author} {\bibfnamefont {S.~R.}\ \bibnamefont {Elliott}}, \bibinfo {author} {\bibfnamefont {R.}~\bibnamefont {Atta-Fynn}}, \ and\ \bibinfo {author} {\bibfnamefont {P.}~\bibnamefont {Biswas}},\ }\bibfield  {title} {\enquote {\bibinfo {title} {Disorder by design: A data-driven approach to amorphous semiconductors without total-energy functionals},}\ }\href@noop {} {\bibfield  {journal} {\bibinfo  {journal} {Sci. Rep.}\ }\textbf {\bibinfo {volume} {10}},\ \bibinfo {pages} {7742} (\bibinfo {year} {2020})}\BibitemShut {NoStop}%
\bibitem [{\citenamefont {Opletal}\ \emph {et~al.}(2017)\citenamefont {Opletal}, \citenamefont {Petersen}, \citenamefont {Barnard},\ and\ \citenamefont {Russo}}]{opletal_2017}%
  \BibitemOpen
  \bibfield  {author} {\bibinfo {author} {\bibfnamefont {G.}~\bibnamefont {Opletal}}, \bibinfo {author} {\bibfnamefont {T.~C.}\ \bibnamefont {Petersen}}, \bibinfo {author} {\bibfnamefont {A.~S.}\ \bibnamefont {Barnard}}, \ and\ \bibinfo {author} {\bibfnamefont {S.~P.}\ \bibnamefont {Russo}},\ }\bibfield  {title} {\enquote {\bibinfo {title} {On reverse {Monte Carlo} constraints and model reproduction},}\ }\href@noop {} {\bibfield  {journal} {\bibinfo  {journal} {J. Comput. Chem.}\ }\textbf {\bibinfo {volume} {38}},\ \bibinfo {pages} {1547} (\bibinfo {year} {2017})}\BibitemShut {NoStop}%
\bibitem [{\citenamefont {Opletal}\ \emph {et~al.}(2013)\citenamefont {Opletal}, \citenamefont {Petersen}, \citenamefont {Snook},\ and\ \citenamefont {Russo}}]{opletal_2013}%
  \BibitemOpen
  \bibfield  {author} {\bibinfo {author} {\bibfnamefont {G.}~\bibnamefont {Opletal}}, \bibinfo {author} {\bibfnamefont {T.~C.}\ \bibnamefont {Petersen}}, \bibinfo {author} {\bibfnamefont {I.~K.}\ \bibnamefont {Snook}}, \ and\ \bibinfo {author} {\bibfnamefont {S.~P.}\ \bibnamefont {Russo}},\ }\bibfield  {title} {\enquote {\bibinfo {title} {{HRMC}\_2.0: Hybrid reverse {Monte Carlo} method with silicon, carbon and germanium potentials},}\ }\href@noop {} {\bibfield  {journal} {\bibinfo  {journal} {Comput. Phys. Comm.}\ }\textbf {\bibinfo {volume} {184}},\ \bibinfo {pages} {1946} (\bibinfo {year} {2013})}\BibitemShut {NoStop}%
\bibitem [{\citenamefont {Khadka}\ \emph {et~al.}(2020)\citenamefont {Khadka}, \citenamefont {Baishnab}, \citenamefont {Opletal},\ and\ \citenamefont {Sakidja}}]{khadka_2020}%
  \BibitemOpen
  \bibfield  {author} {\bibinfo {author} {\bibfnamefont {R.}~\bibnamefont {Khadka}}, \bibinfo {author} {\bibfnamefont {N.}~\bibnamefont {Baishnab}}, \bibinfo {author} {\bibfnamefont {G.}~\bibnamefont {Opletal}}, \ and\ \bibinfo {author} {\bibfnamefont {R.}~\bibnamefont {Sakidja}},\ }\bibfield  {title} {\enquote {\bibinfo {title} {Study of amorphous boron carbide (a-{B$_x$C}) materials using molecular dynamics ({MD}) and hybrid reverse {Monte Carlo} ({HRMC})},}\ }\href@noop {} {\bibfield  {journal} {\bibinfo  {journal} {J. Non-Cryst. Solids}\ }\textbf {\bibinfo {volume} {530}},\ \bibinfo {pages} {119783} (\bibinfo {year} {2020})}\BibitemShut {NoStop}%
\bibitem [{\citenamefont {Farmahini}\ \emph {et~al.}(2013)\citenamefont {Farmahini}, \citenamefont {Opletal},\ and\ \citenamefont {Bhatia}}]{farmahini_2013}%
  \BibitemOpen
  \bibfield  {author} {\bibinfo {author} {\bibfnamefont {A.~H.}\ \bibnamefont {Farmahini}}, \bibinfo {author} {\bibfnamefont {G.}~\bibnamefont {Opletal}}, \ and\ \bibinfo {author} {\bibfnamefont {S.~K.}\ \bibnamefont {Bhatia}},\ }\bibfield  {title} {\enquote {\bibinfo {title} {Structural modelling of silicon carbide-derived nanoporous carbon by hybrid reverse {Monte Carlo} simulation},}\ }\href@noop {} {\bibfield  {journal} {\bibinfo  {journal} {J. Phys. Chem. C}\ }\textbf {\bibinfo {volume} {117}},\ \bibinfo {pages} {14081} (\bibinfo {year} {2013})}\BibitemShut {NoStop}%
\bibitem [{\citenamefont {Cons~Bacilla}\ \emph {et~al.}(2022)\citenamefont {Cons~Bacilla}, \citenamefont {Futamura},\ and\ \citenamefont {Iiyama}}]{bacilla_2022}%
  \BibitemOpen
  \bibfield  {author} {\bibinfo {author} {\bibfnamefont {A.~C.}\ \bibnamefont {Cons~Bacilla}}, \bibinfo {author} {\bibfnamefont {R.}~\bibnamefont {Futamura}}, \ and\ \bibinfo {author} {\bibfnamefont {T.}~\bibnamefont {Iiyama}},\ }\bibfield  {title} {\enquote {\bibinfo {title} {Effect of quadrupole of nitrogen, as a probe molecule for surface area estimation: {XRD} and {HRMC} investigation},}\ }\href@noop {} {\bibfield  {journal} {\bibinfo  {journal} {Chem. Lett.}\ }\textbf {\bibinfo {volume} {51}},\ \bibinfo {pages} {338} (\bibinfo {year} {2022})}\BibitemShut {NoStop}%
\bibitem [{\citenamefont {Nicholas}\ \emph {et~al.}(2024)\citenamefont {Nicholas}, \citenamefont {Stones}, \citenamefont {Patel}, \citenamefont {Michel}, \citenamefont {Reeder}, \citenamefont {Aarts}, \citenamefont {Deringer},\ and\ \citenamefont {Goodwin}}]{nicholas_2024}%
  \BibitemOpen
  \bibfield  {author} {\bibinfo {author} {\bibfnamefont {T.~C.}\ \bibnamefont {Nicholas}}, \bibinfo {author} {\bibfnamefont {A.~E.}\ \bibnamefont {Stones}}, \bibinfo {author} {\bibfnamefont {A.}~\bibnamefont {Patel}}, \bibinfo {author} {\bibfnamefont {F.~M.}\ \bibnamefont {Michel}}, \bibinfo {author} {\bibfnamefont {R.~J.}\ \bibnamefont {Reeder}}, \bibinfo {author} {\bibfnamefont {D.~G.~A.~L.}\ \bibnamefont {Aarts}}, \bibinfo {author} {\bibfnamefont {V.~L.}\ \bibnamefont {Deringer}}, \ and\ \bibinfo {author} {\bibfnamefont {A.~L.}\ \bibnamefont {Goodwin}},\ }\bibfield  {title} {\enquote {\bibinfo {title} {Geometrically frustrated interactions drive structural complexity in amorphous calcium carbonate},}\ }\href@noop {} {\bibfield  {journal} {\bibinfo  {journal} {Nat. Chem.}\ }\textbf {\bibinfo {volume} {16}},\ \bibinfo {pages} {36} (\bibinfo {year} {2024})}\BibitemShut {NoStop}%
\bibitem [{\citenamefont {Pandey}\ \emph {et~al.}(2016)\citenamefont {Pandey}, \citenamefont {Biswas},\ and\ \citenamefont {Drabold}}]{pandey_2016}%
  \BibitemOpen
  \bibfield  {author} {\bibinfo {author} {\bibfnamefont {A.}~\bibnamefont {Pandey}}, \bibinfo {author} {\bibfnamefont {P.}~\bibnamefont {Biswas}}, \ and\ \bibinfo {author} {\bibfnamefont {D.~A.}\ \bibnamefont {Drabold}},\ }\bibfield  {title} {\enquote {\bibinfo {title} {Inversion of diffraction data for amorphous materials},}\ }\href@noop {} {\bibfield  {journal} {\bibinfo  {journal} {Sci. Rep.}\ }\textbf {\bibinfo {volume} {6}},\ \bibinfo {pages} {33731} (\bibinfo {year} {2016})}\BibitemShut {NoStop}%
\bibitem [{\citenamefont {de~Tomas}\ \emph {et~al.}(2019)\citenamefont {de~Tomas}, \citenamefont {Aghajamali}, \citenamefont {Jones}, \citenamefont {Lim}, \citenamefont {L{\'o}pez}, \citenamefont {Suarez-Martinez},\ and\ \citenamefont {Marks}}]{detomas_2019}%
  \BibitemOpen
  \bibfield  {author} {\bibinfo {author} {\bibfnamefont {C.}~\bibnamefont {de~Tomas}}, \bibinfo {author} {\bibfnamefont {A.}~\bibnamefont {Aghajamali}}, \bibinfo {author} {\bibfnamefont {J.~L.}\ \bibnamefont {Jones}}, \bibinfo {author} {\bibfnamefont {D.~J.}\ \bibnamefont {Lim}}, \bibinfo {author} {\bibfnamefont {M.~J.}\ \bibnamefont {L{\'o}pez}}, \bibinfo {author} {\bibfnamefont {I.}~\bibnamefont {Suarez-Martinez}}, \ and\ \bibinfo {author} {\bibfnamefont {N.~A.}\ \bibnamefont {Marks}},\ }\bibfield  {title} {\enquote {\bibinfo {title} {Transferability in interatomic potentials for carbon},}\ }\href@noop {} {\bibfield  {journal} {\bibinfo  {journal} {Carbon}\ }\textbf {\bibinfo {volume} {155}},\ \bibinfo {pages} {624} (\bibinfo {year} {2019})}\BibitemShut {NoStop}%
\bibitem [{\citenamefont {Caro}\ \emph {et~al.}(2020)\citenamefont {Caro}, \citenamefont {Cs{\'a}nyi}, \citenamefont {Laurila},\ and\ \citenamefont {Deringer}}]{caro_2020c}%
  \BibitemOpen
  \bibfield  {author} {\bibinfo {author} {\bibfnamefont {M.~A.}\ \bibnamefont {Caro}}, \bibinfo {author} {\bibfnamefont {G.}~\bibnamefont {Cs{\'a}nyi}}, \bibinfo {author} {\bibfnamefont {T.}~\bibnamefont {Laurila}}, \ and\ \bibinfo {author} {\bibfnamefont {V.~L.}\ \bibnamefont {Deringer}},\ }\bibfield  {title} {\enquote {\bibinfo {title} {Machine learning driven simulated deposition of carbon films: from low-density to diamondlike amorphous carbon},}\ }\href@noop {} {\bibfield  {journal} {\bibinfo  {journal} {Phys. Rev. B}\ }\textbf {\bibinfo {volume} {102}},\ \bibinfo {pages} {174201} (\bibinfo {year} {2020})}\BibitemShut {NoStop}%
\bibitem [{\citenamefont {Marchant}\ \emph {et~al.}(2023)\citenamefont {Marchant}, \citenamefont {Caro}, \citenamefont {Karasulu},\ and\ \citenamefont {P\'artay}}]{marchant_2023}%
  \BibitemOpen
  \bibfield  {author} {\bibinfo {author} {\bibfnamefont {G.}~\bibnamefont {Marchant}}, \bibinfo {author} {\bibfnamefont {M.~A.}\ \bibnamefont {Caro}}, \bibinfo {author} {\bibfnamefont {B.}~\bibnamefont {Karasulu}}, \ and\ \bibinfo {author} {\bibfnamefont {L.~B.}\ \bibnamefont {P\'artay}},\ }\bibfield  {title} {\enquote {\bibinfo {title} {Exploring the configuration space of elemental carbon with empirical and machine learned interatomic potentials},}\ }\href@noop {} {\bibfield  {journal} {\bibinfo  {journal} {npj Comput. Mater.}\ }\textbf {\bibinfo {volume} {9}},\ \bibinfo {pages} {131} (\bibinfo {year} {2023})}\BibitemShut {NoStop}%
\bibitem [{\citenamefont {Perdew}\ \emph {et~al.}(1996)\citenamefont {Perdew}, \citenamefont {Burke},\ and\ \citenamefont {Ernzerhof}}]{perdew_1996}%
  \BibitemOpen
  \bibfield  {author} {\bibinfo {author} {\bibfnamefont {J.~P.}\ \bibnamefont {Perdew}}, \bibinfo {author} {\bibfnamefont {K.}~\bibnamefont {Burke}}, \ and\ \bibinfo {author} {\bibfnamefont {M.}~\bibnamefont {Ernzerhof}},\ }\bibfield  {title} {\enquote {\bibinfo {title} {Generalized gradient approximation made simple},}\ }\href@noop {} {\bibfield  {journal} {\bibinfo  {journal} {Phys. Rev. Lett.}\ }\textbf {\bibinfo {volume} {77}},\ \bibinfo {pages} {3865} (\bibinfo {year} {1996})}\BibitemShut {NoStop}%
\bibitem [{\citenamefont {Kresse}\ and\ \citenamefont {Furthm{\"u}ller}(1996)}]{kresse_1996}%
  \BibitemOpen
  \bibfield  {author} {\bibinfo {author} {\bibfnamefont {G.}~\bibnamefont {Kresse}}\ and\ \bibinfo {author} {\bibfnamefont {J.}~\bibnamefont {Furthm{\"u}ller}},\ }\bibfield  {title} {\enquote {\bibinfo {title} {Efficient iterative schemes for \textit{ab initio} total-energy calculations using a plane-wave basis set},}\ }\href@noop {} {\bibfield  {journal} {\bibinfo  {journal} {Phys. Rev. B}\ }\textbf {\bibinfo {volume} {54}},\ \bibinfo {pages} {11169} (\bibinfo {year} {1996})}\BibitemShut {NoStop}%
\bibitem [{\citenamefont {Kresse}\ and\ \citenamefont {Joubert}(1999)}]{kresse_1999}%
  \BibitemOpen
  \bibfield  {author} {\bibinfo {author} {\bibfnamefont {G.}~\bibnamefont {Kresse}}\ and\ \bibinfo {author} {\bibfnamefont {D.}~\bibnamefont {Joubert}},\ }\bibfield  {title} {\enquote {\bibinfo {title} {From ultrasoft pseudopotentials to the projector augmented-wave method},}\ }\href@noop {} {\bibfield  {journal} {\bibinfo  {journal} {Phys. Rev. B}\ }\textbf {\bibinfo {volume} {59}},\ \bibinfo {pages} {1758} (\bibinfo {year} {1999})}\BibitemShut {NoStop}%
\bibitem [{\citenamefont {Bart{\'o}k}\ and\ \citenamefont {Cs{\'a}nyi}(2015)}]{bartok_2015}%
  \BibitemOpen
  \bibfield  {author} {\bibinfo {author} {\bibfnamefont {A.P.}\ \bibnamefont {Bart{\'o}k}}\ and\ \bibinfo {author} {\bibfnamefont {G.}~\bibnamefont {Cs{\'a}nyi}},\ }\bibfield  {title} {\enquote {\bibinfo {title} {Gaussian approximation potentials: A brief tutorial introduction},}\ }\href@noop {} {\bibfield  {journal} {\bibinfo  {journal} {Int. J. Quantum Chem.}\ }\textbf {\bibinfo {volume} {115}},\ \bibinfo {pages} {1051} (\bibinfo {year} {2015})}\BibitemShut {NoStop}%
\bibitem [{\citenamefont {Musil}\ \emph {et~al.}(2021)\citenamefont {Musil}, \citenamefont {Grisafi}, \citenamefont {Bart{\'o}k}, \citenamefont {Ortner}, \citenamefont {Cs{\'a}nyi},\ and\ \citenamefont {Ceriotti}}]{musil_2021}%
  \BibitemOpen
  \bibfield  {author} {\bibinfo {author} {\bibfnamefont {F.}~\bibnamefont {Musil}}, \bibinfo {author} {\bibfnamefont {A.}~\bibnamefont {Grisafi}}, \bibinfo {author} {\bibfnamefont {A.~P.}\ \bibnamefont {Bart{\'o}k}}, \bibinfo {author} {\bibfnamefont {C.}~\bibnamefont {Ortner}}, \bibinfo {author} {\bibfnamefont {G.}~\bibnamefont {Cs{\'a}nyi}}, \ and\ \bibinfo {author} {\bibfnamefont {M.}~\bibnamefont {Ceriotti}},\ }\bibfield  {title} {\enquote {\bibinfo {title} {Physics-inspired structural representations for molecules and materials},}\ }\href@noop {} {\bibfield  {journal} {\bibinfo  {journal} {Chem. Rev.}\ }\textbf {\bibinfo {volume} {121}},\ \bibinfo {pages} {9759} (\bibinfo {year} {2021})}\BibitemShut {NoStop}%
\bibitem [{\citenamefont {Caro}(2019)}]{caro_2019}%
  \BibitemOpen
  \bibfield  {author} {\bibinfo {author} {\bibfnamefont {M.~A.}\ \bibnamefont {Caro}},\ }\bibfield  {title} {\enquote {\bibinfo {title} {Optimizing many-body atomic descriptors for enhanced computational performance of machine learning based interatomic potentials},}\ }\href@noop {} {\bibfield  {journal} {\bibinfo  {journal} {Phys. Rev. B}\ }\textbf {\bibinfo {volume} {100}},\ \bibinfo {pages} {024112} (\bibinfo {year} {2019})}\BibitemShut {NoStop}%
\bibitem [{\citenamefont {Bart{\'o}k}\ \emph {et~al.}(2013)\citenamefont {Bart{\'o}k}, \citenamefont {Kondor},\ and\ \citenamefont {Cs{\'a}nyi}}]{bartok_2013}%
  \BibitemOpen
  \bibfield  {author} {\bibinfo {author} {\bibfnamefont {A.~P.}\ \bibnamefont {Bart{\'o}k}}, \bibinfo {author} {\bibfnamefont {R.}~\bibnamefont {Kondor}}, \ and\ \bibinfo {author} {\bibfnamefont {G.}~\bibnamefont {Cs{\'a}nyi}},\ }\bibfield  {title} {\enquote {\bibinfo {title} {On representing chemical environments},}\ }\href@noop {} {\bibfield  {journal} {\bibinfo  {journal} {Phys. Rev. B}\ }\textbf {\bibinfo {volume} {87}},\ \bibinfo {pages} {184115} (\bibinfo {year} {2013})}\BibitemShut {NoStop}%
\bibitem [{\citenamefont {Caro}\ and\ \citenamefont {Zarrouk}(2024)}]{caro_2024}%
  \BibitemOpen
  \bibfield  {author} {\bibinfo {author} {\bibfnamefont {M.~A.}\ \bibnamefont {Caro}}\ and\ \bibinfo {author} {\bibfnamefont {T.}~\bibnamefont {Zarrouk}},\ }\bibfield  {title} {\enquote {\bibinfo {title} {General purpose {Gaussian} approximation potential for {CO}},}\ }\href {\doibase DOI:10.5281/zenodo.10456068} {\bibfield  {journal} {\bibinfo  {journal} {Zenodo}\ } (\bibinfo {year} {2024}),\ DOI:10.5281/zenodo.10456068}\BibitemShut {NoStop}%
\bibitem [{ref(accessed: 2024-01-20{\natexlab{a}})}]{ref_turbogap}%
  \BibitemOpen
  \href@noop {} {\enquote {\bibinfo {title} {{TurboGAP}: Data-driven atomistic simulations},}\ }\bibinfo {howpublished} {\url{http://turbogap.fi}} (\bibinfo {year} {accessed: 2024-01-20}{\natexlab{a}})\BibitemShut {NoStop}%
\bibitem [{\citenamefont {Cs{\'a}nyi}\ \emph {et~al.}(2007)\citenamefont {Cs{\'a}nyi}, \citenamefont {Winfield}, \citenamefont {Kermode}, \citenamefont {De~Vita}, \citenamefont {Comisso}, \citenamefont {Bernstein},\ and\ \citenamefont {Payne}}]{csanyi_2007}%
  \BibitemOpen
  \bibfield  {author} {\bibinfo {author} {\bibfnamefont {G.}~\bibnamefont {Cs{\'a}nyi}}, \bibinfo {author} {\bibfnamefont {S.}~\bibnamefont {Winfield}}, \bibinfo {author} {\bibfnamefont {J.~R.}\ \bibnamefont {Kermode}}, \bibinfo {author} {\bibfnamefont {A.}~\bibnamefont {De~Vita}}, \bibinfo {author} {\bibfnamefont {A.}~\bibnamefont {Comisso}}, \bibinfo {author} {\bibfnamefont {N.}~\bibnamefont {Bernstein}}, \ and\ \bibinfo {author} {\bibfnamefont {M.~C.}\ \bibnamefont {Payne}},\ }\bibfield  {title} {\enquote {\bibinfo {title} {Expressive programming for computational physics in {Fortran 95+}},}\ }\href@noop {} {\bibfield  {journal} {\bibinfo  {journal} {IoP Comput. Phys. Newsletter}\ ,\ \bibinfo {pages} {Spring 2007}} (\bibinfo {year} {2007})}\BibitemShut {NoStop}%
\bibitem [{ref(accessed: 2024-01-20{\natexlab{b}})}]{ref_quip}%
  \BibitemOpen
  \href@noop {} {\enquote {\bibinfo {title} {Quip},}\ }\bibinfo {howpublished} {\url{https://libatoms.github.io/QUIP}} (\bibinfo {year} {accessed: 2024-01-20}{\natexlab{b}})\BibitemShut {NoStop}%
\bibitem [{\citenamefont {Plimpton}(1995)}]{plimpton_1995}%
  \BibitemOpen
  \bibfield  {author} {\bibinfo {author} {\bibfnamefont {S.}~\bibnamefont {Plimpton}},\ }\bibfield  {title} {\enquote {\bibinfo {title} {Fast parallel algorithms for short-range molecular dynamics},}\ }\href@noop {} {\bibfield  {journal} {\bibinfo  {journal} {J. Comput. Phys.}\ }\textbf {\bibinfo {volume} {117}},\ \bibinfo {pages} {1} (\bibinfo {year} {1995})}\BibitemShut {NoStop}%
\bibitem [{ref(accessed: 2024-01-20{\natexlab{c}})}]{ref_lammps}%
  \BibitemOpen
  \href@noop {} {\enquote {\bibinfo {title} {Large-scale atomic/molecular massively parallel simulator},}\ }\bibinfo {howpublished} {\url{http://lammps.sandia.gov}} (\bibinfo {year} {accessed: 2024-01-20}{\natexlab{c}})\BibitemShut {NoStop}%
\bibitem [{\citenamefont {Deringer}\ and\ \citenamefont {Cs{\'a}nyi}(2017)}]{deringer_2017}%
  \BibitemOpen
  \bibfield  {author} {\bibinfo {author} {\bibfnamefont {V.~L.}\ \bibnamefont {Deringer}}\ and\ \bibinfo {author} {\bibfnamefont {G.}~\bibnamefont {Cs{\'a}nyi}},\ }\bibfield  {title} {\enquote {\bibinfo {title} {Machine learning based interatomic potential for amorphous carbon},}\ }\href@noop {} {\bibfield  {journal} {\bibinfo  {journal} {Phys. Rev. B}\ }\textbf {\bibinfo {volume} {95}},\ \bibinfo {pages} {094203} (\bibinfo {year} {2017})}\BibitemShut {NoStop}%
\bibitem [{\citenamefont {Sure}\ \emph {et~al.}(2017)\citenamefont {Sure}, \citenamefont {Hansen}, \citenamefont {Schwerdtfeger},\ and\ \citenamefont {Grimme}}]{sure_2017}%
  \BibitemOpen
  \bibfield  {author} {\bibinfo {author} {\bibfnamefont {R.}~\bibnamefont {Sure}}, \bibinfo {author} {\bibfnamefont {A.}~\bibnamefont {Hansen}}, \bibinfo {author} {\bibfnamefont {P.}~\bibnamefont {Schwerdtfeger}}, \ and\ \bibinfo {author} {\bibfnamefont {S.}~\bibnamefont {Grimme}},\ }\bibfield  {title} {\enquote {\bibinfo {title} {Comprehensive theoretical study of all 1812 {C$_{60}$} isomers},}\ }\href@noop {} {\bibfield  {journal} {\bibinfo  {journal} {Phys. Chem. Chem. Phys.}\ }\textbf {\bibinfo {volume} {19}},\ \bibinfo {pages} {14296} (\bibinfo {year} {2017})}\BibitemShut {NoStop}%
\bibitem [{\citenamefont {Caro}(2023)}]{caro_2023}%
  \BibitemOpen
  \bibfield  {author} {\bibinfo {author} {\bibfnamefont {M.~A.}\ \bibnamefont {Caro}},\ }\bibfield  {title} {\enquote {\bibinfo {title} {Machine learning based modeling of disordered elemental semiconductors: understanding the atomic structure of {a-Si} and {a-C}},}\ }\href@noop {} {\bibfield  {journal} {\bibinfo  {journal} {Semicond. Sci. Technol.}\ }\textbf {\bibinfo {volume} {38}},\ \bibinfo {pages} {043001} (\bibinfo {year} {2023})}\BibitemShut {NoStop}%
\bibitem [{\citenamefont {Klawohn}\ \emph {et~al.}(2023{\natexlab{a}})\citenamefont {Klawohn}, \citenamefont {Cs\'anyi}, \citenamefont {Darby}, \citenamefont {Kermode}, \citenamefont {Caro},\ and\ \citenamefont {Bart\'ok}}]{klawohn_2023}%
  \BibitemOpen
  \bibfield  {author} {\bibinfo {author} {\bibfnamefont {S.}~\bibnamefont {Klawohn}}, \bibinfo {author} {\bibfnamefont {G.}~\bibnamefont {Cs\'anyi}}, \bibinfo {author} {\bibfnamefont {J.~P.}\ \bibnamefont {Darby}}, \bibinfo {author} {\bibfnamefont {J.~R.}\ \bibnamefont {Kermode}}, \bibinfo {author} {\bibfnamefont {M.~A.}\ \bibnamefont {Caro}}, \ and\ \bibinfo {author} {\bibfnamefont {A.~P.}\ \bibnamefont {Bart\'ok}},\ }\bibfield  {title} {\enquote {\bibinfo {title} {Gaussian approximation potentials: theory, software implementation and application examples},}\ }\href@noop {} {\bibfield  {journal} {\bibinfo  {journal} {J. Chem. Phys.}\ }\textbf {\bibinfo {volume} {159}},\ \bibinfo {pages} {174108} (\bibinfo {year} {2023}{\natexlab{a}})}\BibitemShut {NoStop}%
\bibitem [{\citenamefont {Klawohn}\ \emph {et~al.}(2023{\natexlab{b}})\citenamefont {Klawohn}, \citenamefont {Kermode},\ and\ \citenamefont {Bart{\'o}k}}]{klawohn_2023b}%
  \BibitemOpen
  \bibfield  {author} {\bibinfo {author} {\bibfnamefont {S.}~\bibnamefont {Klawohn}}, \bibinfo {author} {\bibfnamefont {J.~R.}\ \bibnamefont {Kermode}}, \ and\ \bibinfo {author} {\bibfnamefont {A.~P.}\ \bibnamefont {Bart{\'o}k}},\ }\bibfield  {title} {\enquote {\bibinfo {title} {Massively parallel fitting of {Gaussian} approximation potentials},}\ }\href@noop {} {\bibfield  {journal} {\bibinfo  {journal} {Mach. Learn. Sci. Technol.}\ }\textbf {\bibinfo {volume} {4}},\ \bibinfo {pages} {015020} (\bibinfo {year} {2023}{\natexlab{b}})}\BibitemShut {NoStop}%
\bibitem [{\citenamefont {Larsen}\ \emph {et~al.}(2017)\citenamefont {Larsen}, \citenamefont {Mortensen}, \citenamefont {Blomqvist}, \citenamefont {Castelli}, \citenamefont {Christensen}, \citenamefont {Dulak}, \citenamefont {Friis}, \citenamefont {Groves}, \citenamefont {Hammer}, \citenamefont {Hargus}, \citenamefont {Hermes}, \citenamefont {Jennings}, \citenamefont {Jensen}, \citenamefont {Kermode}, \citenamefont {Kitchin}, \citenamefont {Kolsbjerg}, \citenamefont {Kubal}, \citenamefont {Kaasbjerg}, \citenamefont {Lysgaard}, \citenamefont {Maronsson}, \citenamefont {Maxson}, \citenamefont {Olsen}, \citenamefont {Pastewka}, \citenamefont {Peterson}, \citenamefont {Rostgaard}, \citenamefont {Schi{\o}tz}, \citenamefont {Sch\"utt}, \citenamefont {Strange}, \citenamefont {Thygesen}, \citenamefont {Vegge}, \citenamefont {Vilhelmsen}, \citenamefont {Walter}, \citenamefont {Zeng},\ and\ \citenamefont {Jacobsen}}]{larsen_2017}%
  \BibitemOpen
  \bibfield  {author} {\bibinfo {author} {\bibfnamefont {A.~H.}\ \bibnamefont {Larsen}}, \bibinfo {author} {\bibfnamefont {J.~J.}\ \bibnamefont {Mortensen}}, \bibinfo {author} {\bibfnamefont {J.}~\bibnamefont {Blomqvist}}, \bibinfo {author} {\bibfnamefont {I.~E.}\ \bibnamefont {Castelli}}, \bibinfo {author} {\bibfnamefont {R.}~\bibnamefont {Christensen}}, \bibinfo {author} {\bibfnamefont {M.}~\bibnamefont {Dulak}}, \bibinfo {author} {\bibfnamefont {J.}~\bibnamefont {Friis}}, \bibinfo {author} {\bibfnamefont {M.~N.}\ \bibnamefont {Groves}}, \bibinfo {author} {\bibfnamefont {B.}~\bibnamefont {Hammer}}, \bibinfo {author} {\bibfnamefont {C.}~\bibnamefont {Hargus}}, \bibinfo {author} {\bibfnamefont {E.~D.}\ \bibnamefont {Hermes}}, \bibinfo {author} {\bibfnamefont {P.~C.}\ \bibnamefont {Jennings}}, \bibinfo {author} {\bibfnamefont {P.~B.}\ \bibnamefont {Jensen}}, \bibinfo {author} {\bibfnamefont {J.}~\bibnamefont {Kermode}}, \bibinfo {author} {\bibfnamefont {J.~R.}\ \bibnamefont {Kitchin}}, \bibinfo {author}
  {\bibfnamefont {E.~L.}\ \bibnamefont {Kolsbjerg}}, \bibinfo {author} {\bibfnamefont {J.}~\bibnamefont {Kubal}}, \bibinfo {author} {\bibfnamefont {K.}~\bibnamefont {Kaasbjerg}}, \bibinfo {author} {\bibfnamefont {S.}~\bibnamefont {Lysgaard}}, \bibinfo {author} {\bibfnamefont {J.~B.}\ \bibnamefont {Maronsson}}, \bibinfo {author} {\bibfnamefont {T.}~\bibnamefont {Maxson}}, \bibinfo {author} {\bibfnamefont {T.}~\bibnamefont {Olsen}}, \bibinfo {author} {\bibfnamefont {L.}~\bibnamefont {Pastewka}}, \bibinfo {author} {\bibfnamefont {A.}~\bibnamefont {Peterson}}, \bibinfo {author} {\bibfnamefont {C.}~\bibnamefont {Rostgaard}}, \bibinfo {author} {\bibfnamefont {J.}~\bibnamefont {Schi{\o}tz}}, \bibinfo {author} {\bibfnamefont {O.}~\bibnamefont {Sch\"utt}}, \bibinfo {author} {\bibfnamefont {M.}~\bibnamefont {Strange}}, \bibinfo {author} {\bibfnamefont {K.~S.}\ \bibnamefont {Thygesen}}, \bibinfo {author} {\bibfnamefont {T.}~\bibnamefont {Vegge}}, \bibinfo {author} {\bibfnamefont {L.}~\bibnamefont {Vilhelmsen}}, \bibinfo
  {author} {\bibfnamefont {M.}~\bibnamefont {Walter}}, \bibinfo {author} {\bibfnamefont {Z.}~\bibnamefont {Zeng}}, \ and\ \bibinfo {author} {\bibfnamefont {K.~W.}\ \bibnamefont {Jacobsen}},\ }\bibfield  {title} {\enquote {\bibinfo {title} {The {Atomic Simulation Environment} -- {A Python} library for working with atoms},}\ }\href@noop {} {\bibfield  {journal} {\bibinfo  {journal} {J. Phys.: Condens. Matter}\ }\textbf {\bibinfo {volume} {29}},\ \bibinfo {pages} {273002} (\bibinfo {year} {2017})}\BibitemShut {NoStop}%
\bibitem [{\citenamefont {Frenkel}\ and\ \citenamefont {Smit}(2002)}]{frenkel_2002}%
  \BibitemOpen
  \bibfield  {author} {\bibinfo {author} {\bibfnamefont {D.}~\bibnamefont {Frenkel}}\ and\ \bibinfo {author} {\bibfnamefont {B.}~\bibnamefont {Smit}},\ }\href@noop {} {\emph {\bibinfo {title} {Understanding Molecular Simulation: From Algorithms to Applications}}},\ \bibinfo {edition} {2nd}\ ed.\ (\bibinfo  {publisher} {Academic Press},\ \bibinfo {address} {San Diego},\ \bibinfo {year} {2002})\BibitemShut {NoStop}%
\bibitem [{\citenamefont {Hagberg}\ \emph {et~al.}(2008)\citenamefont {Hagberg}, \citenamefont {Schult},\ and\ \citenamefont {Swart}}]{SciPyProceedings_11}%
  \BibitemOpen
  \bibfield  {author} {\bibinfo {author} {\bibfnamefont {A.~A.}\ \bibnamefont {Hagberg}}, \bibinfo {author} {\bibfnamefont {D.~A.}\ \bibnamefont {Schult}}, \ and\ \bibinfo {author} {\bibfnamefont {P.~J.}\ \bibnamefont {Swart}},\ }\bibfield  {title} {\enquote {\bibinfo {title} {Exploring network structure, dynamics, and function using {{NetworkX}}},}\ }in\ \href@noop {} {\emph {\bibinfo {booktitle} {Proceedings of the 7th Python in Science Conference}}},\ \bibinfo {editor} {edited by\ \bibinfo {editor} {\bibfnamefont {G.}~\bibnamefont {Varoquaux}}, \bibinfo {editor} {\bibfnamefont {T.}~\bibnamefont {Vaught}}, \ and\ \bibinfo {editor} {\bibfnamefont {J.}~\bibnamefont {Millman}}}\ (\bibinfo {address} {{Pasadena, CA USA}},\ \bibinfo {year} {2008})\ p.~\bibinfo {pages} {11}\BibitemShut {NoStop}%
\bibitem [{\citenamefont {Marks}(2005)}]{marks_2005}%
  \BibitemOpen
  \bibfield  {author} {\bibinfo {author} {\bibfnamefont {N.~A.}\ \bibnamefont {Marks}},\ }\bibfield  {title} {\enquote {\bibinfo {title} {Thin film deposition of tetrahedral amorphous carbon: a molecular dynamics study},}\ }\href@noop {} {\bibfield  {journal} {\bibinfo  {journal} {Diam. Relat. Mater.}\ }\textbf {\bibinfo {volume} {14}},\ \bibinfo {pages} {1223} (\bibinfo {year} {2005})}\BibitemShut {NoStop}%
\bibitem [{\citenamefont {Caro}\ \emph {et~al.}(2018{\natexlab{b}})\citenamefont {Caro}, \citenamefont {Deringer}, \citenamefont {Koskinen}, \citenamefont {Laurila},\ and\ \citenamefont {Cs\'anyi}}]{caro_2018}%
  \BibitemOpen
  \bibfield  {author} {\bibinfo {author} {\bibfnamefont {M.~A.}\ \bibnamefont {Caro}}, \bibinfo {author} {\bibfnamefont {V.~L.}\ \bibnamefont {Deringer}}, \bibinfo {author} {\bibfnamefont {J.}~\bibnamefont {Koskinen}}, \bibinfo {author} {\bibfnamefont {T.}~\bibnamefont {Laurila}}, \ and\ \bibinfo {author} {\bibfnamefont {G.}~\bibnamefont {Cs\'anyi}},\ }\bibfield  {title} {\enquote {\bibinfo {title} {Growth mechanism and origin of high $sp^3$ content in tetrahedral amorphous carbon},}\ }\href@noop {} {\bibfield  {journal} {\bibinfo  {journal} {Phys. Rev. Lett.}\ }\textbf {\bibinfo {volume} {120}},\ \bibinfo {pages} {166101} (\bibinfo {year} {2018}{\natexlab{b}})}\BibitemShut {NoStop}%
\bibitem [{\citenamefont {Marks}\ \emph {et~al.}(1996)\citenamefont {Marks}, \citenamefont {McKenzie}, \citenamefont {Pailthorpe}, \citenamefont {Bernasconi},\ and\ \citenamefont {Parrinello}}]{marks_1996}%
  \BibitemOpen
  \bibfield  {author} {\bibinfo {author} {\bibfnamefont {N.~A.}\ \bibnamefont {Marks}}, \bibinfo {author} {\bibfnamefont {D.~R.}\ \bibnamefont {McKenzie}}, \bibinfo {author} {\bibfnamefont {B.~A.}\ \bibnamefont {Pailthorpe}}, \bibinfo {author} {\bibfnamefont {M.}~\bibnamefont {Bernasconi}}, \ and\ \bibinfo {author} {\bibfnamefont {M.}~\bibnamefont {Parrinello}},\ }\bibfield  {title} {\enquote {\bibinfo {title} {\textit{Ab initio} simulations of tetrahedral amorphous carbon},}\ }\href@noop {} {\bibfield  {journal} {\bibinfo  {journal} {Phys. Rev. B}\ }\textbf {\bibinfo {volume} {54}},\ \bibinfo {pages} {9703} (\bibinfo {year} {1996})}\BibitemShut {NoStop}%
\bibitem [{\citenamefont {Caro}\ \emph {et~al.}(2014)\citenamefont {Caro}, \citenamefont {Zoubkoff}, \citenamefont {Lopez-Acevedo},\ and\ \citenamefont {Laurila}}]{caro_2014}%
  \BibitemOpen
  \bibfield  {author} {\bibinfo {author} {\bibfnamefont {M.~A.}\ \bibnamefont {Caro}}, \bibinfo {author} {\bibfnamefont {R.}~\bibnamefont {Zoubkoff}}, \bibinfo {author} {\bibfnamefont {O.}~\bibnamefont {Lopez-Acevedo}}, \ and\ \bibinfo {author} {\bibfnamefont {T.}~\bibnamefont {Laurila}},\ }\bibfield  {title} {\enquote {\bibinfo {title} {Atomic and electronic structure of tetrahedral amorphous carbon surfaces from density functional theory: Properties and simulation strategies},}\ }\href@noop {} {\bibfield  {journal} {\bibinfo  {journal} {Carbon}\ }\textbf {\bibinfo {volume} {77}},\ \bibinfo {pages} {1168} (\bibinfo {year} {2014})}\BibitemShut {NoStop}%
\bibitem [{\citenamefont {Laurila}\ \emph {et~al.}(2017)\citenamefont {Laurila}, \citenamefont {Sainio},\ and\ \citenamefont {Caro}}]{laurila_2017}%
  \BibitemOpen
  \bibfield  {author} {\bibinfo {author} {\bibfnamefont {T.}~\bibnamefont {Laurila}}, \bibinfo {author} {\bibfnamefont {S.}~\bibnamefont {Sainio}}, \ and\ \bibinfo {author} {\bibfnamefont {M.~A.}\ \bibnamefont {Caro}},\ }\bibfield  {title} {\enquote {\bibinfo {title} {Hybrid carbon based nanomaterials for electrochemical detection of biomolecules},}\ }\href@noop {} {\bibfield  {journal} {\bibinfo  {journal} {Prog. Mater. Sci.}\ }\textbf {\bibinfo {volume} {88}},\ \bibinfo {pages} {499} (\bibinfo {year} {2017})}\BibitemShut {NoStop}%
\bibitem [{\citenamefont {de~Tomas}\ \emph {et~al.}(2016)\citenamefont {de~Tomas}, \citenamefont {Suarez-Martinez},\ and\ \citenamefont {Marks}}]{detomas_2016}%
  \BibitemOpen
  \bibfield  {author} {\bibinfo {author} {\bibfnamefont {C.}~\bibnamefont {de~Tomas}}, \bibinfo {author} {\bibfnamefont {I.}~\bibnamefont {Suarez-Martinez}}, \ and\ \bibinfo {author} {\bibfnamefont {N.~A.}\ \bibnamefont {Marks}},\ }\bibfield  {title} {\enquote {\bibinfo {title} {Graphitization of amorphous carbons: A comparative study of interatomic potentials},}\ }\href@noop {} {\bibfield  {journal} {\bibinfo  {journal} {Carbon}\ }\textbf {\bibinfo {volume} {109}},\ \bibinfo {pages} {681} (\bibinfo {year} {2016})}\BibitemShut {NoStop}%
\bibitem [{\citenamefont {Jana}\ \emph {et~al.}(2019)\citenamefont {Jana}, \citenamefont {Savio}, \citenamefont {Deringer},\ and\ \citenamefont {Pastewka}}]{jana_2019}%
  \BibitemOpen
  \bibfield  {author} {\bibinfo {author} {\bibfnamefont {R.}~\bibnamefont {Jana}}, \bibinfo {author} {\bibfnamefont {D.}~\bibnamefont {Savio}}, \bibinfo {author} {\bibfnamefont {V.~L.}\ \bibnamefont {Deringer}}, \ and\ \bibinfo {author} {\bibfnamefont {L.}~\bibnamefont {Pastewka}},\ }\bibfield  {title} {\enquote {\bibinfo {title} {Structural and elastic properties of amorphous carbon from simulated quenching at low rates},}\ }\href@noop {} {\bibfield  {journal} {\bibinfo  {journal} {Model. Simul. Mater. Sc.}\ }\textbf {\bibinfo {volume} {27}},\ \bibinfo {pages} {085009} (\bibinfo {year} {2019})}\BibitemShut {NoStop}%
\bibitem [{\citenamefont {Wang}\ \emph {et~al.}(2022)\citenamefont {Wang}, \citenamefont {Fan}, \citenamefont {Qian}, \citenamefont {Ala-Nissila},\ and\ \citenamefont {Caro}}]{wang_2022}%
  \BibitemOpen
  \bibfield  {author} {\bibinfo {author} {\bibfnamefont {Y.}~\bibnamefont {Wang}}, \bibinfo {author} {\bibfnamefont {Z.}~\bibnamefont {Fan}}, \bibinfo {author} {\bibfnamefont {P.}~\bibnamefont {Qian}}, \bibinfo {author} {\bibfnamefont {T.}~\bibnamefont {Ala-Nissila}}, \ and\ \bibinfo {author} {\bibfnamefont {M.~A.}\ \bibnamefont {Caro}},\ }\bibfield  {title} {\enquote {\bibinfo {title} {Structure and pore size distribution in nanoporous carbon},}\ }\href@noop {} {\bibfield  {journal} {\bibinfo  {journal} {Chem. Mater.}\ }\textbf {\bibinfo {volume} {34}},\ \bibinfo {pages} {617} (\bibinfo {year} {2022})}\BibitemShut {NoStop}%
\bibitem [{\citenamefont {Samanta}\ \emph {et~al.}(2010)\citenamefont {Samanta}, \citenamefont {Lenosky},\ and\ \citenamefont {Li}}]{samanta_2010}%
  \BibitemOpen
  \bibfield  {author} {\bibinfo {author} {\bibfnamefont {A.}~\bibnamefont {Samanta}}, \bibinfo {author} {\bibfnamefont {T.}~\bibnamefont {Lenosky}}, \ and\ \bibinfo {author} {\bibfnamefont {J.}~\bibnamefont {Li}},\ }\bibfield  {title} {\enquote {\bibinfo {title} {Thermodynamic stability of oxygen point defects in cubic zirconia},}\ }\href@noop {} {\bibfield  {journal} {\bibinfo  {journal} {arXiv:1009.5567}\ } (\bibinfo {year} {2010})}\BibitemShut {NoStop}%
\bibitem [{\citenamefont {Haubner}\ \emph {et~al.}(2010)\citenamefont {Haubner}, \citenamefont {Murawski}, \citenamefont {Olk}, \citenamefont {Eng}, \citenamefont {Ziegler}, \citenamefont {Adolphi},\ and\ \citenamefont {Jaehne}}]{Haubner2010}%
  \BibitemOpen
  \bibfield  {author} {\bibinfo {author} {\bibfnamefont {K.}~\bibnamefont {Haubner}}, \bibinfo {author} {\bibfnamefont {J.}~\bibnamefont {Murawski}}, \bibinfo {author} {\bibfnamefont {P.}~\bibnamefont {Olk}}, \bibinfo {author} {\bibfnamefont {L.~M.}\ \bibnamefont {Eng}}, \bibinfo {author} {\bibfnamefont {C.}~\bibnamefont {Ziegler}}, \bibinfo {author} {\bibfnamefont {B.}~\bibnamefont {Adolphi}}, \ and\ \bibinfo {author} {\bibfnamefont {E.}~\bibnamefont {Jaehne}},\ }\bibfield  {title} {\enquote {\bibinfo {title} {The {{Route}} to {{Functional Graphene Oxide}}},}\ }\href {\doibase 10.1002/cphc.201000132} {\bibfield  {journal} {\bibinfo  {journal} {ChemPhysChem}\ }\textbf {\bibinfo {volume} {11}},\ \bibinfo {pages} {2131} (\bibinfo {year} {2010})}\BibitemShut {NoStop}%
\bibitem [{\citenamefont {Oganov}\ and\ \citenamefont {Glass}(2006)}]{Oganov2006}%
  \BibitemOpen
  \bibfield  {author} {\bibinfo {author} {\bibfnamefont {Artem~R.}\ \bibnamefont {Oganov}}\ and\ \bibinfo {author} {\bibfnamefont {Colin~W.}\ \bibnamefont {Glass}},\ }\bibfield  {title} {\enquote {\bibinfo {title} {Crystal structure prediction using {\emph{ab initio}} evolutionary techniques: {{Principles}} and applications},}\ }\href {\doibase 10.1063/1.2210932} {\bibfield  {journal} {\bibinfo  {journal} {The Journal of Chemical Physics}\ }\textbf {\bibinfo {volume} {124}},\ \bibinfo {pages} {244704} (\bibinfo {year} {2006})}\BibitemShut {NoStop}%
\bibitem [{\citenamefont {Oganov}\ \emph {et~al.}(2011)\citenamefont {Oganov}, \citenamefont {Lyakhov},\ and\ \citenamefont {Valle}}]{Oganov2011}%
  \BibitemOpen
  \bibfield  {author} {\bibinfo {author} {\bibfnamefont {Artem~R.}\ \bibnamefont {Oganov}}, \bibinfo {author} {\bibfnamefont {Andriy~O.}\ \bibnamefont {Lyakhov}}, \ and\ \bibinfo {author} {\bibfnamefont {Mario}\ \bibnamefont {Valle}},\ }\bibfield  {title} {\enquote {\bibinfo {title} {How {{Evolutionary Crystal Structure Prediction Works}}---and {{Why}}},}\ }\href {\doibase 10.1021/ar1001318} {\bibfield  {journal} {\bibinfo  {journal} {Accounts of Chemical Research}\ }\textbf {\bibinfo {volume} {44}},\ \bibinfo {pages} {227--237} (\bibinfo {year} {2011})}\BibitemShut {NoStop}%
\bibitem [{\citenamefont {Lyakhov}\ \emph {et~al.}(2013)\citenamefont {Lyakhov}, \citenamefont {Oganov}, \citenamefont {Stokes},\ and\ \citenamefont {Zhu}}]{Lyakhov2013}%
  \BibitemOpen
  \bibfield  {author} {\bibinfo {author} {\bibfnamefont {Andriy~O.}\ \bibnamefont {Lyakhov}}, \bibinfo {author} {\bibfnamefont {Artem~R.}\ \bibnamefont {Oganov}}, \bibinfo {author} {\bibfnamefont {Harold~T.}\ \bibnamefont {Stokes}}, \ and\ \bibinfo {author} {\bibfnamefont {Qiang}\ \bibnamefont {Zhu}},\ }\bibfield  {title} {\enquote {\bibinfo {title} {New developments in evolutionary structure prediction algorithm {{USPEX}}},}\ }\href {\doibase 10.1016/j.cpc.2012.12.009} {\bibfield  {journal} {\bibinfo  {journal} {Computer Physics Communications}\ }\textbf {\bibinfo {volume} {184}},\ \bibinfo {pages} {1172--1182} (\bibinfo {year} {2013})}\BibitemShut {NoStop}%
\end{thebibliography}
\end{document}